\documentstyle[12pt]{article} 
{\baselineskip   12pt} 
\begin{document} 
\begin{flushright} {\bf  WSP--IF  97--50}  \\ 
{\bf July 1997} 
\end{flushright} 
\hfill \\ 
\begin{center} {\Large \bf CPT transformation properties} \\ 
\hfill \\ 
{\Large  \bf  of  the  exact effective Hamiltonian }\\ 
\hfill \\ 
{\Large \bf   for  neutral  kaon and similar complexes.} \\ 
\hfill \\  \hfill  \\  K.Urbanowski  \\  \hfill  \\ Pedagogical 
University, Institute  of  Physics,  \\ 
Plac Slowianski 6, 65-069 Zielona Gora, Poland. \\  \hfill  \\  July 
21, 1997 
\end{center} 
\hfill  \\ 
{\noindent}PACS  numbers: 03.65.Bz., 11.30.Er.,  13.20.Eb. \\
\nopagebreak 
\begin{abstract} 
CPT--symmetry  
properties  of   the  exact  effective  Hamiltonian  $H_{\parallel}$ 
governing the time evolution in  the ${\rm  K}_{0}$,  $\overline{\rm  
K}_{0}$   mesons   subspace implied by such properties of the  total 
Hamiltonian   $H$   of    the  system   under   consideration    are   
examined.   We    show    that  $H_{\parallel}$  can  commute   with  
${\cal C}{\cal P}{\cal T}$-- operator only if $H$ does  not  commute 
with it. We also find  that,  in    contradistinction     to     the   
standard    result    of    the   Lee--Oehme--Yang   (LOY)   theory, 
Re$<K_{0}|H_{\parallel}|K_{0}> =$ 
Re$<\overline{K}_{0}|H_{\parallel}|\overline{K}_{0}>$,         i.e., 
$(<K_{0}|H_{\parallel}|K_{0}>  -  <\overline{K}_{0}|   H_{\parallel} 
|\overline{K}_{0}>) = 0$,  only   if   the  total  system  does  not 
preserve  CPT--symmetry.  Using  more  accurate  approximation  than 
Weisskopf--Wigner approximation, an  estimation  of  the  difference 
$(<K_{0}|H_{\parallel}|K_{0}>          -          <\overline{K}_{0}| 
H_{\parallel}|\overline{K}_{0}>)$  is   found   for   CPT--invariant 
generalized  Fridrichs--Lee  model. 
\end{abstract} 
\pagebreak[4] 
\section{Introduction.} 
The  problem  of  testing   CPT--invariance 
experimentally  has   attracted   the   attention   of   physicsist, 
practically since the discovery of antiparticles. CPT symmetry is  a 
fundamental theorem of axiomatic quantum field theory which  follows 
from locality, Lorentz invariance, and  unitarity  \cite{cpt}.  Many 
tests of CPT--invariance consist in searching for decay  process  of 
neutral kaons. All known CP--  and  hypothetically   possible  CPT-- 
violation  effects in neutral kaon complex are  described by solving 
the   Schr\"{o}dinger--like   evolution   equation   \cite{1}    --- 
\cite{chiu} (we use $\hbar  =  c  =  1$  units) 
\begin{equation} 
i 
\frac{\partial}{\partial t} |\psi ; t >_{\parallel} =  H_{\parallel} 
|\psi ; t >_{\parallel} \label{l1} 
\end{equation}  for  $|\psi  ;  t >_{\parallel}$  belonging  to  the 
subspace  ${\cal  H}_{\parallel} 
\subset {\cal H}$ (where ${\cal  H}$  is  the  state  space  of  the 
physical system under investigation), e.g., spanned  by  orthonormal 
neutral  kaons states $|K_{0}>, \; |{\overline{K}}_{0}>$, and so on, 
(then states corresponding with the decay products belong to  ${\cal 
H}  \ominus  {\cal  H}_{\parallel}   \stackrel{\rm   def}{=}   {\cal 
H}_{\perp}$), and nonhermitean effective Hamiltonian $H_{\parallel}$ 
obtained usually by means of  the so--called  LOY approach   (within 
the   Weisskopf--Wigner approximation (WW)) \cite{1}  ---  \cite{4}: 
\begin{equation}  H_{\parallel}  \equiv  M  -  \frac{i}{2}   \Gamma, 
\label{new1} \end{equation} where \begin{equation} M = M^{+}, \;  \; 
\Gamma = {\Gamma}^{+}, \label{new1a} \end{equation} are  $(2  \times 
2)$ matrices.

Solutions of Eq. (\ref{l1}) can be written  in  matrix  
form  and such  a  matrix  defines  the evolution    operator (which    
is     usually     nonunitary) $U_{\parallel}(t)$ acting  in  ${\cal 
H}_{\parallel}$:  \begin{equation}  |\psi  ;   t   >_{\parallel}   = 
U_{\parallel}(t)  |\psi  ;t_{0}  =  0  >_{\parallel}   \stackrel{\rm 
def}{=}   U_{\parallel}(t)    |\psi    >_{\parallel},    \label{l1a} 
\end{equation} where, \begin{equation}  |\psi  >_{\parallel}  \equiv 
q_{1}|{\bf 1}> +  q_{2}|{\bf  2}>,  \label{l1b}  \end{equation}  and 
$|{\bf 1}>$ stands for  the  vectors of the   $|K_{0}>,  \; |B_{0}>$ 
type and $|{\bf 2}>$ denotes antiparticles  of  the   particle  "1": 
$|{\overline{K}}_{0}>, \; {\overline{B}}_{0}>$, $<{\bf j}|{\bf k}> = 
{\delta}_{jk}$, $j,k =1,2$.

Relations  between  matrix  elements  of  
$H_{\parallel}$  implied  by CPT--transformation properties  of  the 
Hamiltonian $H$ of  the  total  system,  containing   neutral   kaon  
complex   as   a   subsystem,   are  crucial  for  designing   CPT-- 
invariance and CP--violation tests  and for proper interpretation of 
their  results.  (Main  properties  of  the  effective   Hamiltonian 
$H_{\parallel}$ and formulae appearing in the LOY approach  will  be 
described in short in Section 2).  The  aim  of  this  paper  is  to 
examine the properties of the exact $H_{\parallel}$ generated by the 
CPT--symmetry  of  the  total   system   under   consideration   and 
independent of the approximation used.  In  order  to  realize  this 
purpose, the method decribed and applied to study the properties  of 
time evolution in neutral kaon system in \cite{8} --- \cite{10} and, 
especially, in \cite{11}, will be used. For the readers  convenience 
this   method   will   be   sketched   briefly   in    Section    3.

\section{Preliminaries.} 
\subsection{Properties  of  eigenstates  of $H_{\parallel}$.} 

The eigenstates of  $H_{\parallel}$,   $|l>$   and  
$|s>$,    for    the   eigenvalues   ${\mu}_{l}$   and   ${\mu}_{s}$ 
respectively  \cite{1}  ---  \cite{chiu},  \cite{8}  ---   \cite{10} 
\begin{equation} {\mu}_{l(s)} = h_{0} -  (+)  h  \equiv  m_{l(s)}  - 
\frac{i}{2}  {\gamma}_{l(s)},    \label{r5}   \end{equation}   where 
$m_{l(s)}, {\gamma}_{l(s)}$ are real, and \begin{eqnarray} h_{0} & = 
& \frac{1}{2}(h_{11} + h_{22}), \label{r5a} \\ h & \equiv  &  \sqrt{ 
h_{z}^{2} + h_{12} h_{21} }, \label{r5b} \\ h_{z} & = &  \frac{1}{2} 
(h_{11}  -  h_{22}),   \label{r5c}   \\   h_{jk}   &   =   &   <{\bf 
j}|H_{\parallel}|{\bf k}>, \; (j,k=1,2),  \end{eqnarray}  correspond 
to the long (the vector $|l>$) and  short (the vector $|s>$)  living 
superpositions of $K_{0}$  and   $\overline{K_{0}}$.

The  following 
identity taking place for ${\mu}_{l}$ and ${\mu}_{s}$ will be needed 
in next Sections: \begin{eqnarray}  {\mu}_{l}  +  {\mu}_{s}  &  =  & 
h_{11} + h_{22} , \label{new5}  \\ {\mu}_{s} - {\mu}_{l} &  =  &  2h 
\stackrel{\rm def}{=}  \Delta  \mu,  \label{new6}  \\  {\mu}_{l}  \, 
{\mu}_{s}  &  =  &  h_{11}  h_{22}  -   h_{12}h_{21},   \label{new7} 
\end{eqnarray}

Using the  eigenvectors  \begin{equation}  |K_{1(2)}> 
\stackrel{\rm def}{=}  2^{-1/2}  (|{\bf  1}>  +  (-)  |{\bf  2}>)  , 
\label{new8}  \end{equation}  of  the  CP--transformation  for   the 
eigenvalues $\pm  1$ (we define ${\cal C}{\cal  P}  |{\bf  1}>  =  - 
|{\bf 2}>$, \linebreak ${\cal C}{\cal P} | {\bf 2}> = - |{\bf 1}>$), 
vectors  $|l>$ and $|s>$ can be expressed  as  follows  \cite{3,4,5} 
\begin{equation} |l(s)> \equiv (1  +  |{\varepsilon}_{l(s)}|^{2})^{- 
1/2} [|K_{2(1)} > + {\varepsilon}_{l(s)} |K_{1(2)} > ] ,  \label{r6} 
\end{equation}  where  \begin{eqnarray}  {\varepsilon}_{l}  &  =   & 
\frac{h_{12} - h_{11} +  {\mu}_{l}}{h_{12}  +  h_{11}  -  {\mu}_{l}} 
\equiv - \frac{h_{21}  -  h_{22}  +  {\mu}_{l}}{h_{21}  +  h_{22}  - 
{\mu}_{l}}, \label{new9} \\ {\varepsilon}_{s} & = &  \frac{h_{12}  + 
h_{11}  -  {\mu}_{s}}{h_{12}  -  h_{11}  +   {\mu}_{l}}   \equiv   - 
\frac{h_{21} + h_{22} - {\mu}_{s}}{h_{21}  -  h_{22}  +  {\mu}_{s}}, 
\label{new10} \end{eqnarray} This form of $|l>$ and $|s>$ is used in 
many papers when possible departures from CP-- or  CPT--symmetry  in  
the system considered are discussed. The  following  parameters  are 
used to describe the scale  of CP-- and possible  CPT  --  violation 
effects  \cite{4,5}:  \begin{equation}   \varepsilon   \stackrel{\rm 
def}{=} \frac{1}{2} (  {\varepsilon}_{s}  +  {\varepsilon}_{l}  )  , 
\label{r7}  \end{equation}  \begin{equation}  \delta   \stackrel{\rm 
def}{=} \frac{1}{2} (  {\varepsilon}_{s}  -  {\varepsilon}_{l}  )  . 
\label{r8} \end{equation} According  to   the   standard    meaning,   
$\varepsilon$ describes violations of CP--symmetry and  $\delta$  is 
considered as  a  CPT--violating  parameter  \cite{2,4,5}.  Such  an 
interpretation of  these  parameters  follows from properties of LOY 
theory of  time   evolution  in  the  subspace   of  neutral   kaons  
\cite{1}. We have 
\begin{eqnarray} \varepsilon & = & \frac{h_{12}  - 
h_{21}}{D} \label{r9a} \\ \delta & = & \frac{h_{11}  -  h_{22}  }{D} 
\equiv  \frac{2  h_{z}  }{D}  ,  \label{r9}   \end{eqnarray}   where 
\begin{equation} D  \stackrel{\rm def}{=}  h_{12} + h_{21} +  \Delta 
\mu . \label{new11} \end{equation}

Starting from  Eqs.  (\ref{new5}) 
--- (\ref{new7}) and  (\ref{new9}),  (\ref{new10})  and  using  some 
known identities for ${\mu}_{l}, {\mu}_{s}$  one can express  matrix 
elements $h_{jk}$  of  $H_{\parallel}$  in  terms  of  the  physical 
parameters      ${\varepsilon}_{l(s)}$      and      ${\mu}_{l(s)}$: 
\begin{eqnarray} h_{11} & =  &  \frac{{\mu}_{s}  +  {\mu}_{l}}{2}  + 
\frac{{\mu}_{s}   -    {\mu}_{l}}{2}    \frac{{\varepsilon}_{s}    - 
{\varepsilon}_{l}}{ 1  -  {\varepsilon}_{l}  {\varepsilon}_{s}  }  , 
\label{new12} \\ h_{22} & =  &  \frac{{\mu}_{s}  +  {\mu}_{l}}{2}  - 
\frac{{\mu}_{s}   -    {\mu}_{l}}{2}    \frac{{\varepsilon}_{s}    - 
{\varepsilon}_{l}}{ 1  -  {\varepsilon}_{l}  {\varepsilon}_{s}  }  , 
\label{new13} \\  h_{12}  &  =  &  \frac{{\mu}_{s}  -  {\mu}_{l}}{2} 
\frac{(1  +  {\varepsilon}_{l})(1  +  {\varepsilon}_{s})}  {   1   - 
{\varepsilon}_{l}{\varepsilon}_{s} } , \label{new14} \\ h_{21} & = & 
\frac{{\mu}_{s} - {\mu}_{l}}{2} \frac{(1  -  {\varepsilon}_{l})(1  - 
{\varepsilon}_{s})} { 1  -  {\varepsilon}_{l}{\varepsilon}_{s}  }  . 
\label{new15} \end{eqnarray} These relations lead to  the  following 
equations  \begin{eqnarray}  h_{11}  -  h_{22}  &  =  &  \Delta  \mu 
\frac{{\varepsilon}_{s} - {\varepsilon}_{l}}{ 1 -  {\varepsilon}_{l} 
{\varepsilon}_{s} } , \label{new16} \\ h_{12} + h_{21} & = &  \Delta 
\mu   \frac{1   +{\varepsilon}_{l}   {\varepsilon}_{s}   }{   1    - 
{\varepsilon}_{l} {\varepsilon}_{s} } , \label{new17}  \\  h_{12}  - 
h_{21}   &   =    &    \Delta    \mu    \frac{{\varepsilon}_{s}    + 
{\varepsilon}_{l}}{ 1  -  {\varepsilon}_{l}  {\varepsilon}_{s}  }  , 
\label{new18} \end{eqnarray} Note that relations  (\ref{new12})  --- 
(\ref{new18})    are    valid    for     arbitrary     values     of 
${\varepsilon}_{l(s)}$.  From  (\ref{new16})  one  infers  that   if 
$\Delta \mu \neq  0$  then:  \begin{equation}  h_{11}  =  h_{22}  \; 
\Longleftrightarrow   \;{\varepsilon}_{l}    =    {\varepsilon}_{s}. 
\label{new19} \end{equation} Relation (\ref{new18})  enables  us  to 
conclude     that       parametres      ${\varepsilon}_{l}$      and 
${\varepsilon}_{s}$ need not be small, in order  $\varepsilon  =  0$ 
(\ref{r9a}). Indeed, the identity  (\ref{new18})  implies  that  for 
$\Delta  \mu  \neq  0$   \begin{equation}   h_{12}   =   h_{21}   \; 
\Longleftrightarrow  \;  {\varepsilon}_{l}  =  -  {\varepsilon}_{s}, 
\label{new20} \end{equation} for any values of $|{\varepsilon}_{l}|, 
|{\varepsilon}_{s}|$.

It is appropriate to emphasize at  this  point 
that all relations (\ref{new12}) --- (\ref{new20}) do not depend  on 
a special form of the effective  Hamiltonian  $H_{\parallel}$.  They 
are induced by geometric relations between various base  vectors  in 
two--dimensional subspace ${\cal H}_{\parallel}$. On the other hand, 
the interpretation of above relations depends on properties  of  the 
matrix   elements   $h_{jk}$   of    the    effective    Hamiltonian 
$H_{\parallel}$, i.e., if for example $H_{\parallel} \neq  H_{LOY}$, 
where  $H_{LOY}$  is  the  LOY  effective  Hamiltonian,   then   the 
interpretation of $\varepsilon$ (\ref{r7}) and  $\delta$  (\ref{r8}) 
etc., need not be the same for $H_{\parallel}$  and  for  $H_{LOY}$.

Experimentally measured  values  of  parameters  ${\varepsilon}_{l}, 
{\varepsilon}_{s}$  are  very  small  for  neutral  kaons.  Assuming 
\begin{equation}    |{\varepsilon}_{l}|     \ll     1,     \;     \; 
|{\varepsilon}_{s}|  \ll  1,   \label{new21}   \end{equation}   from 
({\ref{new16}) one finds: \begin{equation} h_{11}  -  h_{22}  \simeq 
({\mu}_{s} - {\mu}_{l}) ({\varepsilon}_{s}  -  {\varepsilon}_{l}  ), 
\label{new22}    \end{equation}    and    ({\ref{new17})     implies 
\begin{equation} h_{12}  +  h_{21}  \simeq  {\mu}_{s}  -  {\mu}_{l}, 
\label{new23}     \end{equation}     and     (\ref{new18})     gives 
\begin{equation} h_{12} -  h_{21}  \simeq  ({\mu}_{s}  -  {\mu}_{l}) 
({\varepsilon}_{s}     +      {\varepsilon}_{l}).      \label{new24} 
\end{equation} Relation (\ref{new23}) means that in  the  considered 
case   of   small   values   of   parameters   $|{\varepsilon}_{l}|, 
|{\varepsilon}_{s}|$ (\ref{new21}), the quantity  $D$  (\ref{new11}) 
appearing in formulae for $\delta$ and  $\varepsilon$  approximately 
equals \begin{equation} D \simeq 2({\mu}_{s} - {\mu}_{l})  \equiv  2 
\Delta \mu .  \label{new25}  \end{equation}

Keeping  in  mind  that 
$h_{jk}  =  M_{jk}  -   \frac{i}{2}   {\Gamma}_{jk}   ,   M_{kj}   = 
M_{jk}^{\ast},{\Gamma}_{kj}   =   {\Gamma}_{jk}^{\ast}$   and   then 
starting form Eqs. (\ref{new22}) ---  (\ref{new24})  and  separating 
real and  imaginary  parts  one  can  find  some  useful  relations: 
\begin{eqnarray} 2{\rm Re.} (M_{12})  &  \simeq  &  m_{s}  -  m_{l}, 
\label{new26} \\ 2{\rm Re}({\Gamma}_{12}) & \simeq & {\gamma}_{s}  - 
{\gamma}_{l}, \label{new27}  \\  2{\rm  Im}(M_{12})  &  \simeq  &  - 
({\gamma}_{s} - {\gamma}_{l}) [ {\rm Im} (\frac{{\varepsilon}_{s}  + 
{\varepsilon}_{l}}{2})      +      \tan       {\phi}_{SW}       {\rm 
Re}(\frac{{\varepsilon}_{s}   +   {\varepsilon}_{l}}{2}   )   ]    , 
\label{new28} \\ {\rm Im}({\Gamma}_{12}) & \simeq & -  ({\gamma}_{s} 
- {\gamma}_{l}) [ \tan {\phi}_{SW} {\rm Re} (\frac{{\varepsilon}_{s} 
+ {\varepsilon}_{l}}{2})   -  {\rm  Im}  (\frac{{\varepsilon}_{s}  + 
{\varepsilon}_{l}}{2})  ] , \label{new29} \end{eqnarray} etc., where 
\begin{equation}    \tan    {\phi}_{SW}    \stackrel{\rm     def}{=} 
\frac{2(m_{l} - m_{s})}{{\gamma}_{s} - {\gamma}_{l}} . \label{new30} 
\end{equation} and \begin{eqnarray}  {\rm  Re}(h_{11}  -  h_{22})  & 
\equiv & M_{11} - M_{22} \nonumber \\ & \simeq & -  ({\gamma}_{s}  - 
{\gamma}_{l}) [ \tan {\phi}_{SW} {\rm Re} (\frac{{\varepsilon}_{s} - 
{\varepsilon}_{l})}{2})   \nonumber   \\   &    \;    &    -    {\rm 
Im}(\frac{{\varepsilon}_{s}     -     {\varepsilon}_{l})}{2})     ], 
\label{new31} \\ - {\rm Im}(h_{11} - h_{22} ) & \equiv & \frac{1}{2} 
({\Gamma}_{11}  -  {\Gamma}_{22}  )  \nonumber   \\   &   \simeq   & 
({\gamma}_{s} - {\gamma}_{l}) [ {\rm Re} (\frac{{\varepsilon}_{s}  - 
{\varepsilon}_{l})}{2}) \nonumber \\ & \; & + \tan {\phi}_{SW}  {\rm 
Im}(\frac{{\varepsilon}_{s}     -     {\varepsilon}_{l})}{2})     ]. 
\label{new32}  \end{eqnarray}  etc..  One   should   remember   that 
relations  (\ref{new26})  ---   (\ref{new29})   and   (\ref{new31}), 
(\ref{new32}) are  valid  only  if  condition  (\ref{new21})  holds. 
Completing the system of these last six relations  one  can  rewrite 
Eq. (\ref{new5}) to obtain \begin{eqnarray} M_{11} + M_{22}  &  =  & 
m_{l} + m_{s}, \label{new33} \\ {\Gamma}_{11} + {\Gamma}_{22} & =  & 
{\gamma}_{l} + {\gamma}_{s}  .  \label{new34}  \end{eqnarray}  These 
last two Equations are exact independently of whether the  condition 
(\ref{new21})  holds  or  not.

\subsection{$H_{LOY}$   and   CPT--symmetry.} 
Now, let us consider briefly some properties of  the  LOY 
model. Let $H$ be total (selfadjoint) Hamiltonian, acting  in  $\cal 
H$   ---  then   the   total  unitary  evolution   operator   $U(t)$  
fulfills    the    Schr\"{o}dinger   equation   \begin{equation}   i 
\frac{\partial}{\partial t} U(t)|\phi > = H U(t)|\phi >,  \; \; U(0) 
= I, \label{l2} \end{equation} where $I$ is  the  unit  operator  in 
$\cal H$, $|\phi > \equiv |\phi ; t_{0} = 0> \in {\cal H}$  is   the  
initial  state  of  the system: \begin{equation}  |\phi   >   \equiv  
|\psi   >_{\parallel}   \label{l2a}  \end{equation}  in   our   case  
$|\phi ;t> = U(t) |\phi >$. Let $P$ denote the  projection  operator 
onto the subspace ${\cal H}_{\parallel}$: \begin{equation} P{\cal H} 
= {\cal H}_{\parallel}, \; \; \; P =  P^{2}  =  P^{+},  \label{new2} 
\end{equation}  then  the  subspace   of   decay   products   ${\cal 
H}_{\perp}$ equals \begin{equation} {\cal  H}_{\perp}   =  (I  -  P) 
{\cal H} \stackrel{\rm def}{=} Q {\cal H}, \; \; \; Q \equiv I -  P. 
\label{l7c} \end{equation} For  the   case  of  neutral   kaons   or  
neutral  $B$--mesons,  etc.,  the projector $P$  can  be  chosen  as 
follows: \begin{equation}  P  \equiv  |{\bf  1}><{\bf  1}|  +  |{\bf 
2}><{\bf  2}|.  \label{l3}  \end{equation}  We  assume   that   time 
independent basis vectors $|K_{0}>$ and  $|{\overline{K}}_{0}>$  are 
defined analogously to corresponding vectors used in LOY  theory  of 
time evolution in neutral kaon  complex  \cite{1}:  Vectors$|K_{0}>$ 
and $|{\overline{K}}_{0}>$  can be identified with  eigenvectors  of 
the so--called free Hamiltonian $H^{(0)} \equiv  H_{strong}  =  H  - 
H_{W}$, where $H_{W}$ denotes weak interaction which is  responsible 
for transitions between eigenvectors of  $H^{(0)}$,  i.e.,  for  the 
decay process. 

In the LOY approach it is assumed that vectors $|{\bf 
1}>$,  $|{\bf 2}>$ considered above are eigenstates of $H^{(0)}$ for 
2-fold degenerate eigenvalue $m_{0}$: \begin{equation} H^{(0)} |{\bf 
j} > = m_{0} |{\bf j }>, \; \;  j = 1,2 . \label{b1}  \end{equation} 
This means that \begin{equation}  [P,  H^{(0)}]  =  0.  \label{new3} 
\end{equation}  

The  condition   guaranteeing   the   occurence   of 
transitions  between  subspaces ${\cal  H}_{\parallel}$  and  ${\cal 
H}_{\perp}$,  i.e.,   a   decay  process   of   states   in   ${\cal 
H}_{\parallel}$,  can   be   written  as  follows   \begin{equation} 
[P,H_{W}] \neq 0 . \label{r32} \end{equation} 

Usually,  in  LOY  and 
related   approaches,   it   is   assumed   that    \begin{equation} 
{\Theta}H^{(0)}{\Theta}^{-1}  =  {H^{(0)}}^{+}  \equiv   H^{(0)}   , 
\label{r31}  \end{equation}  where  $\Theta$  is   the   antiunitary 
operator:  \begin{equation}  \Theta  \stackrel{\rm   def}{=}   {\cal 
C}{\cal P}{\cal T}. \label{new4} \end{equation} 

Relation (\ref{r31}) 
is  a  particular  form   of   the   general   transformation   rule 
\cite{messiah,bohm,cpt,6}: \begin{equation} {\Theta} L {\Theta}^{-1} 
\stackrel{\rm def}{=} L_{CPT}^{+}, \label{9ab} \end{equation}  where 
$L$ is an arbitrary linear  operator.  Basic  properties  of  anti-- 
linear and linear operators,  their  products  and  commutators  are 
described, eg., in  \cite{messiah,bohm,cpt,6}.  Generally,  defining 
the  commutator  of  anti--linear  operator  ${\Theta}$  and  linear 
operators $L, L_{1}, L_{2}$ , appearing, e.g., in formulae discussed 
in  Sec.4,  we  follow   \cite{messiah,bohm}:   \begin{equation}   [ 
{\Theta}, L ]   \equiv   {\Theta}  L  -  L  {\Theta}  ,  \label{l9a} 
\end{equation} where $L$ can be  replaced  by  product  $L  =  L_{1} 
L_{2}$, etc.. For such defined  commutators  all  basic  commutation 
rules hold (see \cite{messiah}, Chap. XV, \S 3 --- \S  5)  including 
the following one: \begin{equation} [ {\Theta}, L_{1} L_{2}]   =   [ 
{\Theta}  ,L_{1}]L_{2}  +  L_{1}  [  {\Theta}  ,L_{2}].  \label{l9b} 
\end{equation} \nopagebreak[4]
On the other hand, to minimalize risk  of  confusion, 
using definitions (\ref{l9a}) and relations of type (\ref{l9b})  one 
should not forget that properties of products  of  anti--linear  and 
linear operators of types  ${\Theta}  L_{1}  L_{2}$,  $L_{1}{\Theta}  
L_{2} $, etc.,  and  (\ref{l9b}),  differ  from  properties  of  the 
transformation rule (\ref{9ab}) which means that in a  general  case 
${\Theta} L - L_{CPT}^{+} {\Theta} = 0$  (but  does  not  mean  that 
$[{\Theta},L] = 0$), and which implies that \[ {\Theta} L_{1}  L_{2} 
{\Theta}^{-1} \equiv ({\Theta} L_{2} {\Theta}^{-1}) ( {\Theta} L_{1} 
{\Theta}^{-1}  ).  \]  

The  subspace   of   neutral   kaons   ${\cal 
H}_{\parallel}$  is  assumed  to  be   invariant   under   $\Theta$: 
\begin{equation}  {\Theta}  P  {\Theta}^{-1}  =  P^{+}   \equiv   P. 
\label{9aa}  \end{equation}

In  the  kaon  rest  frame,  the   time 
evolution is governed by the  Schr\"{o}dinger  equation  (\ref{l2}), 
where the initial state of the  system  has  the  form  (\ref{l2a}), 
(\ref{l1b}).  Within  assumptions  (\ref{b1})  ---  (\ref{r32})  the  
Weisskopf--Wigner  approach  leads  to  the  following  formula  for 
$H_{LOY}$  (e.g., see \cite{1,2,3,5}):  \begin{eqnarray}  H_{LOY}  = 
m_{0} P + PH_{W} P - \Sigma (m_{0}) & \equiv & PHP - \Sigma (m_{0}), 
\label{b3} \\ & = & M_{LOY} - \frac{i}{2}{\Gamma}_{LOY}  \label{b3a} 
\end{eqnarray} where \begin{equation} \Sigma  (  \epsilon  )  =  PHQ 
\frac{1}{QHQ - \epsilon - i 0} QHP.   \label{r24} \end{equation} The 
matrix elements $h_{jk}^{LOY}$  of  $H_{LOY}$  are  \begin{eqnarray} 
h_{jk}^{LOY} & = &  H_{jk} - {\Sigma}_{jk} (m_{0} ) , \; \; \;  (j,k 
=  1,2)  ,   \label{b5}  \\  &  =  &  M_{jk}^{LOY}   -   \frac{i}{2} 
{\Gamma}_{jk}^{LOY} \label{b5a} \end{eqnarray} where, in this  case, 
\begin{equation} H_{jk} = <{\bf j} |H| {\bf k}  >  \equiv  <{\bf  j} 
|(H^{(0)} + H_{W} )| {\bf k} > \equiv m_{0}  {\delta}_{jk}  +  <{\bf 
j}|H_{W}|{\bf k}> , \label{b6} \end{equation} and  ${\Sigma}_{jk}  ( 
\epsilon ) = < {\bf j} \mid \Sigma ( \epsilon )  \mid  {\bf  k}  >$.

Now, if ${\Theta}H_{W}{\Theta}^{-1} = H_{W}^{+} \equiv H_{W}$,  then 
using, e.g., the following phase convention  \cite{2}  ---  \cite{6} 
\begin{equation} \Theta |{\bf 1}> \stackrel{\rm def}{=} - |{\bf 2}>, 
\;\; \Theta|{\bf 2}> \stackrel{\rm def}{=} - |{\bf 1}>, \label{cpt1} 
\end{equation} and taking into account that $< \psi |  \varphi  >  = 
<{\Theta}{\varphi}|{\Theta}{\psi}>$,   one   easily    finds    from 
(\ref{b3})  --  (\ref{b6})   that  \begin{equation}  h_{11}^{LOY}  - 
h_{22}^{LOY} = 0  \label{b8} \end{equation}  in  the  CPT--invariant 
system. This  is  the standard result of the LOY approach  and  this 
is the picture  which one meets in  the  literature   \cite{1}   ---  
\cite{chiu}.  Property  (\ref{b8})  leads  to  the  conclusion  that 
\begin{equation} \delta \simeq {\delta}^{LOY} \equiv 0 .  \label{b9} 
\end{equation}

\section{Properties of the  exact  $H_{\parallel}$  } 

Eq. (\ref{l2}) means that $U(t) = \exp (-itH)$. Now, knowing $U(t)$, 
the evolution  operator  $U_{\parallel}(t)$ (\ref{l1a})  for  ${\cal 
H}_{\parallel}$ can be expressed   using   the   projector  $P$   as 
follows \begin{equation} U_{\parallel}(t) \equiv PU(t)P.  \label{l4} 
\end{equation} We  have  $U_{\parallel}(0)  =  I_{\parallel}  \equiv  
P$,    where  $I_{\parallel}$  is  the  unit  operator   in   ${\cal 
H}_{\parallel}$. This relation must  be   fulfilled   by   solutions 
$U_{\parallel}(t)$ of Eq.(\ref{l1}) with any effective  Hamiltonian, 
and therefore with the exact $H_{\parallel}$ too. In  \cite{horwitz} 
an observation has been made that for  every  effective  Hamiltonian 
$H_{\parallel}$  governing  the time evolution  in  subspace  ${\cal 
H}_{\parallel}$ $\equiv P {\cal H}$, which in general   can   depend 
on  time   $t$   \cite{8}  ---  \cite{11},  \cite{7}  the  following 
identity    holds:     \begin{equation}     H_{\parallel}     \equiv 
H_{\parallel}(t)                       \equiv                      i 
\frac{{\partial}U_{\parallel}(t)}{\partial t}  [U_{\parallel}(t)]^{- 
1}   .   \label{l5}   \end{equation}   We   define   the    operator 
$[U_{\parallel}(t)]^{-1}    $    as     follows     \begin{equation} 
[U_{\parallel}(t)]^{-1}    U_{\parallel}(t)    =    U_{\parallel}(t) 
[U_{\parallel}(t)]^{-1}\equiv  P.  \label{l6}  \end{equation}   (One 
finds that $PU^{-1}(t)P \equiv PU^{+}(t)P = [U_{\parallel}(t)]^{-1}$  
if and only if $[P, U(t)]$ $= 0$).  From  this  definition  one  can 
infer  that  \begin{equation}   P   [U_{\parallel}(t)]^{-1}   \equiv 
[U_{\parallel}(t)]^{-1}   P.   \label{l6a}    \end{equation}    This 
observation toghether with the property (\ref{l4})  means  that  the 
identity  (\ref{l5})  can  be  replaced  by   the   following   one: 
\begin{equation} H_{\parallel}  \equiv  H_{\parallel}(t)   \equiv  i 
\frac{{\partial}U_{\parallel}(t)}{\partial t}  [U_{\parallel}(t)]^{- 
1} P. \label{l5a} \end{equation} It can  be  easily  verified   that  
$H_{\parallel}  \equiv  H_{LOY}$, fulfils the identities  (\ref{l5}) 
and (\ref{l5a}).

A density matrix approach for a description of time 
evolution in $K_{0}, {\overline K}_{0}$ complex  is  sometimes  used 
\cite{12,13}. One can show that the effective Hamiltonian  appearing 
and used in such an approach fulfils the same identity   (\ref{l5}).

Let us notice  that  definition  (\ref{l6})  means   that  operators 
$[U_{\parallel}(t)]^{-1}$,  as  well   as   $U_{\parallel}(t)$   and 
$H_{\parallel}(t)$ act not only in ${\cal H}_{\parallel}$  but  also  
in  $\cal H$.  It is obvious that  all  these  operators  and  their 
products exist in the  case  of   one  dimensional  subspace  ${\cal 
H}_{\parallel}$.  In  the   case    of    two   dimensional   ${\cal 
H}_{\parallel}$, in which we are interested, assuming that  $\cal H$ 
is spanned by a complete set of orthonormal vectors \linebreak ${\{} 
|e_{l}> {\}}_{l \in \Upsilon}$ (where the  cardinality  of  the  set 
$\Upsilon$ of indexes $l \in \Upsilon$ is equal to the dimension  of 
$\cal H$): $<e_{l}|e_{m}> = {\delta}_{mn}$, eg., by eigenvectors for 
$H^{(0)}$ introduced above, and $|e_{1}> \equiv |{\bf 1}>$, $|e_{2}>  
\equiv  |{\bf  2}>$  we have \begin{eqnarray}  U_{\parallel}(t)  &=&  
\sum_{j,k=1}^{2}  u_{jk}  |{\bf j}><{\bf k}|,   \label{d1a}  \\  {[} 
U_{\parallel} (t)  {]}^{-1}  &=&  \sum_{j,k=1}^{2}  {\tilde  u}_{jk} 
|{\bf   j}><{\bf   k}|,   \label{d1b}   \\   H_{\parallel}(t)    &=&  
\sum_{j,k=1}^{2}  h_{jk}(t)  |{\bf  j}><{\bf  k}|,   \label{d1c}  \\ 
{\cal  O}  &=&  \sum_{l,m  \in  \Upsilon}   O_{lm}   |e_{l}><e_{m}|, 
\nonumber \end{eqnarray} where, $\cal O$ is  an  arbitrary  operator 
acting  in  $\cal  H$,  \begin{eqnarray}  u_{jk}  &   =    &   <{\bf 
j}|U_{\parallel}  (t)|{\bf  k}>,  \nonumber  \\  &  \equiv  &  <{\bf 
j}|U(t)|{\bf k}>, \; \; (j,k =1,2), \nonumber \\ {\tilde u}_{jk} & = 
& <{\bf j}|[U_{\parallel}(t)]^{-1} |{\bf  k}>,  \;  \;  (j,k  =1,2), 
\label{d2} \\ h_{jk}(t) & =  & <{\bf j}|H_{\parallel} (t)|{\bf  k}>, 
\; \; (j,k =1,2), \nonumber \\ O_{lm} & = & <e_{l}|{\cal  O}|e_{m}>, 
\nonumber \end{eqnarray} and \begin{eqnarray} d {\tilde u}_{jk} &  = 
& - u_{jk}, \; ({\scriptstyle j \neq k ;  \; j,k =1,2)} ,  \nonumber 
\\ d {\tilde u}_{11} & = &  u_{22}, \label{d3} \\ d {\tilde  u}_{22} 
& = &  u_{11}, \nonumber \\ d \equiv d(t)  = \det U_{\parallel}  (t) 
&  =  &  u_{11}(t)  u_{22}(t)  -  u_{12}(t)  u_{21}(t).   \label{d4} 
\end{eqnarray} This  realization  means  that  all  operators:  $P$, 
$U_{\parallel}$, $U_{\parallel}^{-1}$,  $H_{\parallel}$,  $\cal  O$, 
etc., are isomorphic with (${\cal  N}   \times  {\cal  N}$)  "block" 
matrices ${\rm {\bf M} }$: \begin{equation} {\rm {\bf M}  }=  \left( 
\begin{array}{cc} {\rm {\bf  M}}_{11} & {\rm {\bf M}}_{12}  \\  {\rm 
{\bf  M}}_{21}  &  {\rm  {\bf   M}}_{22}   \end{array}   \right)   ,  
\label{mat} \end{equation}  where  ${\rm  \bf  M}_{11}$  is  (${\cal 
N}_{1}  \times  {\cal  N}_{1}$)  submatrix,   ${\rm   \bf   M}_{12}$ 
represents (${\cal N}_{1} \times {\cal N}_{2}$) submatrix, ${\rm \bf 
M}_{22}$  is  formed  by  (${\cal  N}_{2}  \times   {\cal   N}_{2}$) 
submatrix, etc., and ${\cal N}_{1} +  {\cal N}_{2} = {\cal N}  $,$\, 
\, {\cal N}_{1} = {\dim}  {\cal  H}_{\parallel}$,  $\,  {\cal  N}  = 
{\dim} {\cal H}$  (in  our  case ${\cal N}_{1} = 2$, $\, {\cal N}  = 
\infty$). ${\rm \bf M}_{11} \neq 0$ and $ {\rm \bf M}_{12} = 0, {\rm 
\bf M}_{21} = 0$ and ${\rm \bf M}_{22} = 0$ in the case of operators 
$P$,$U_{\parallel}$, $U_{\parallel}^{-1}$ and $H_{\parallel}$,  but, 
e.g., for  the  evolution  operator  $U(t)$  one  finds  ${\rm  {\bf 
M}}_{jk}  \neq  0,\;  \;  (j,k  =1,2)$.  So,   products   of   type: 
$H_{\parallel}  U_{\parallel}$,   ${\cal   O}   U_{\parallel}^{-1}$, 
$U_{\parallel}^{-1} {\cal  O}$,  etc.,  exist  not  only  in  ${\cal 
H}_{\parallel}$ but also in ${\cal H}$, and  they  are well  defined  
in  $\cal H$.

Relations (\ref{l6}) and (\ref{d1b})  define  uniquely 
the operator  $U_{\parallel}^{-1}(t)$  in  the  considered  case  of 
neutral kaons. Namely,  properties  of  determinants  enable  us  to 
rewrite (\ref{d4})  as follows \begin{eqnarray}  \det  U_{\parallel} 
(t) & \equiv & \det  [{\cal  D}^{-1}  U_{\parallel}  (t)  {\cal  D}] 
\nonumber \\ & = & \det U_{\parallel}^{d} (t) \equiv  {\zeta}_{-}(t) 
\, {\zeta}_{+} (t),  \label{d5} \end{eqnarray} where $\cal D$ is the 
matrix diagonalizing $U_{\parallel}  (t)$,  $U_{\parallel}^{d}  (t)$ 
denotes the diagonal form of $U_{\parallel}(t)$ and ${\zeta}_{-}(t), 
\, {\zeta}_{+} (t)$ are eigenvalues for  the  matrix  $U_{\parallel} 
(t)$. In \cite{11a} the eigenvalue problem for the operator of  type 
(\ref{l4}) has been discussed and  ${\zeta}_{-}(t),  \,  {\zeta}_{+} 
(t)$ have been given as functions of the matrix elements $u_{kl}(t)$ 
of the total evolution operator $U(t) = \exp (-itH)$.  These  matrix 
elements possess the property: $u_{kl}(t)_{0 < t < \infty}  \neq  0$ 
in the  basis  of  eigenvectors  of  $H^{(0)}$.  From  formulae  for 
${\zeta}_{-}(t), \, {\zeta}_{+} (t)$  and  from  the  properties  of 
$u_{kl}(t)$ \cite{11a} it follows  that  ${\zeta}_{-}(t)_{0  \leq  t 
<\infty}\neq 0$ and ${\zeta}_{+} (t)_{0  \leq  t  \infty}  \neq  0$, 
which appears  quite  natural  and  obvious  for  the  neutral  kaon 
complex. Therefore we can conclude that in our case \begin{equation} 
\det U_{\parallel} (t)  \equiv  {\zeta}_{-}(t)  \,  {\zeta}_{+}  (t) 
\neq 0, \label{d6} \end{equation}  which  means  that  the  operator 
$U_{\parallel}^{-1}(t)$  exists  and  is  defined  uniquely  in  the 
subspace ${\cal H}_{\parallel}$ considered by  relations  (\ref{l6}) 
and  (\ref{d1b}).

Generally  the  operator  $U_{\parallel}^{-1}(t)$ 
defined by (\ref{l6}) exists and is unique, and  formula  (\ref{l5}) 
holds  in  subspaces   ${\cal   H}_{\parallel}$   of   $\dim   {\cal 
H}_{\parallel}  \geq  2$  when  the  effective  evolution   operator 
$U_{\parallel}(t)$ (\ref{l4}) for the subsystem considered has  only 
non--zero eigenvalues. Theory presented and discussed below  can  be 
extended  for   all   such   systems.

Discussing   properties   of 
$U_{\parallel}^{-1}(t)$   one   should   remember   that   relations 
(\ref{l6}) and (\ref{d1b}) define this operator uniquely only in the 
subspace ${\cal H}_{\parallel}$. Its extension  (\ref{d1b})  to  the 
whole  state  space  $\cal  H$  is  not  unique.  The  most  general  
"inverse" operator (let us denote  it  by  ${\cal  U}_{\parallel}^{- 
1}(t)$)    fulfiling   relations    (\ref{l6})    has    the    form 
\begin{equation}  {\cal  U}_{\parallel}^{-1}(t)  =  U_{\parallel}^{- 
1}(t) + QO(t)Q, \label{d7}  \end{equation}  where  $U_{\parallel}^{- 
1}(t)$ is given by formula (\ref{d1b}), and $O(t)$ is  an  arbitrary 
operator. Inserting  (\ref{d7})  instead  of  (\ref{d1b})  into  the 
identity (\ref{l5}) does not  change  anything:  Properties  of  the 
operator  $H_{\parallel}(t)$ (\ref{l5}) are the  same  independently 
of whether  the  operator  (\ref{d1b})  or  (\ref{d7})  is  used  to 
calculate $H_{\parallel}(t)$. Therefore searching for transformation 
properties of the  exact  effective  Hamiltonian  for  neutral  kaon 
complex  one  need  not  worry  about  all  possible  extensions  of 
$U_{\parallel}^{-1}(t)$ to the whole $\cal H$. In order to  describe 
correctly properties of $H_{\parallel}(t)$  it  is  sufficient  that 
$U_{\parallel}^{-1}(t)$ is uniquely defined in the  subspace  ${\cal 
H}_{\parallel}$.

In the nontrivial case \begin{equation} [P,H]  \neq 
0, \label{l7} \end{equation} from (\ref{l5a}), using (\ref{l2})  and 
(\ref{l4}) we find  \begin{eqnarray}  H_{\parallel}(t)  &  \equiv  &  
PHU(t)P[U_{\parallel}(t)]^{-1} P \label{l7a1}  \\ & \equiv &  PHP  + 
PHQ U(t) [ U_{\parallel}(t)]^{-1}P \label{l7a}  \\  &  \stackrel{\rm 
def}{=}  &  PHP  +  V_{\parallel}(t).   \label{l7b}   \end{eqnarray} 
(Assumption (\ref{l7})  means   that   transitions  of  states  from 
${\cal H}_{\parallel}$ into  ${\cal  H}_{\perp}$   and  from  ${\cal 
H}_{\perp}$   into   ${\cal   H}_{\parallel}$,   i.e.,   decay   and 
regeneration processes, are allowed). Thus \cite{8}  ---  \cite{10}, 
\cite{7}  \begin{equation}  H_{\parallel}(0)  \equiv  PHP,   \;   \; 
V_{\parallel}(0) = 0, \; \; V_{\parallel} (t \rightarrow  0)  \simeq 
-itPHQHP,   \label{l8}    \end{equation}    so,      in      general   
$H_{\parallel}(0)      \neq$    $H_{\parallel}(t    \gg    t_{0}=0)$ 
\cite{7,8,9} and $V_{\parallel}(t \neq 0)  \neq  V_{\parallel}^{+}(t 
\neq 0)$, $H_{\parallel}(t \neq  0)  \neq  H_{\parallel}^{+}(t  \neq 
0)$.

According to the ideas of the standard scattering  theory,  one  
can  state   that   operator  $H_{\parallel}(t  \rightarrow  \infty) 
\equiv   H_{\parallel}(\infty)$    describes    the    bounded    or 
quasistationary states  of  the  subsystem  considered  and  in this  
sense  it corresponds to   $H_{LOY}$.

\section{CPT  transformations and the exact $H_{\parallel}$. } 

Now let   us   pass   on   to   the 
investigation      of      CPT--transformation     properties     of 
$H_{\parallel}$. In this Section the following assumptions are used: 
The total Hamiltonian $H$ of the system  considered  is  selfadjoint 
(which means that the total evolution operator $U(t)$ is the unitary 
operator and it solves the Schr\"{o}dinger equation (\ref{l2})  $\,$ 
). Orthonormal  basis  vectors  $|{\bf  1}>,  |{\bf  2}>$  are  time 
independent and they are not eigenvectors for the  $H$,  i.e.,  that 
for the projector $P$ (\ref{l3})  the  condition  (\ref{l7})  holds. 
(This  assumption  means  that  stationary  states   will   not   be 
considered). It is also assumed that vectors $|{\bf  1}>,|{\bf  2}>$ 
are related to each other through the  transformation  (\ref{cpt1}). 
Besides there is only one assumption  for  theanti--linear  operator 
${\Theta}$ (\ref{new4}) describing CPT--transformation in $\cal  H$. 
We require CPT--invariance of  ${\cal  H}_{\parallel}$.  This  means 
that  for  projector  $P$  defining  this  subspace   the   relation 
(\ref{9aa})  must  hold.  Due  to  the   property   $P   =   P^{+}$,  
Eq.(\ref{9aa})  can  be   replaced   by   the   following   relation 
\begin{equation}[ \Theta ,P] = 0.  \label{l9}  \end{equation}

Using 
assumption (\ref{l9}) and  the   identity   (\ref{l7a1}), after some 
algebra, one  finds  \cite{14}  (see  Appendix  A)  \begin{equation} 
[\Theta, H_{\parallel}(t)] = {\cal A}(t) + {\cal B}(t),  \label{l10} 
\end{equation} where: \begin{eqnarray} {\cal A}(t) \; &  =  &  \;  P  
[{\Theta},H]  U(t)  P  \bigl(   U_{\parallel}(t)   {\bigr)}^{-1}   , 
\label{l11} \\ {\cal B}(t) \; & = & \;  \Big{\{} PH  -  PH  U(t)   P 
\bigl( U_{\parallel}(t) {\bigr)}^{-1}P \Big{\}} [{\Theta}  ,U(t)]  P 
\bigl( U_{\parallel}(t) {\bigr)}^{-1} \label{l12}  \\ & \equiv &   { 
\Big{\{} } PH  \: -  \:  H_{\parallel}(t) P { \Big{\}}  }  [{\Theta} 
,U(t)]  P  \bigl(  U_{\parallel}(t){\bigr)}^{-1}  \label{l13}  \\  & 
\equiv &  { \Big{\{} } PHQ  \: -  \:  V_{\parallel}(t) P {  \Big{\}} 
}   [{\Theta}   ,U(t)]    P   \bigl(   U_{\parallel}(t){\bigr)}^{-1} 
\label{l14} \end{eqnarray}

We  observe  that  ${\cal  A}(0)  \equiv 
P[\Theta,H]P$ and ${\cal  B}(0)  \equiv  0$.  From  definitions  and 
general properties of operators  $\cal  C$,$\cal  P$  and  $\cal  T$ 
\cite{5,6,messiah,bohm,15} it is known that ${\cal  T}U(t{\neq}0)  = 
U_{T}^{+}(t{\neq}0){\cal T}$ $\neq$  $U(t{\neq}0){\cal T}$ (Wigner's 
definition for $\cal T$ is used), and thereby ${\Theta}U(t \neq 0) = 
U_{CPT}^{+}(t   \neq    0){\Theta}$    \cite{messiah,bohm,15}   i.e. 
$[\Theta,U(t  \neq  0)]  \neq  0$.   So,    the    component  ${\cal 
B}(t)$ in (\ref{l10}) is nonzero for $t \neq 0$ and  it  is  obvious 
that there is a chance for $\Theta$--operator to  commute  with  the 
effective   Hamiltonian   $H_{\parallel}(t   \neq      0)$  only  if  
$[\Theta,H]  \neq  0$.  On   the   other    hand,    the    property 
$[\Theta,H]\neq 0$ does not imply that $[\Theta, H_{\parallel}(0)] =  
0$  or $[\Theta, H_{\parallel}(0)] \neq 0$. These two  possibilities  
are admissible, but if $[\Theta, H] =  0$  then there  is  only  one 
possibility:  $[\Theta,  H_{\parallel}(0)]  =  0$  \cite{11}.

From (\ref{l10})  we  find   
\begin{equation}   \Theta   H_{\parallel}(t) 
\Theta^{-1} - H_{\parallel}(t) \equiv \bigl(  {\cal  A}(t)  +  {\cal 
B}(t) \bigr) \Theta^{-1}. \label{l16} \end{equation} Now, keeping in 
mind that $|{\bf 2}> \equiv  |\overline{K}_{0}>$ is the antiparticle 
for $|{\bf 1}> \equiv |K_{0}>$ and that,  by definition, the  (anti- 
-unitary) $\Theta$--operator  transforms $|{\bf  1}>$    in   $|{\bf 
2}>$   \cite{2} --- \cite{6} according to formulae (\ref{cpt1}), and 
$< \psi | \varphi > = <{\Theta}{\varphi}|{\Theta}{\psi}>$, we obtain 
from(\ref{l16}) (see Appendix A) \begin{equation} h_{11}(t)^{\ast} - 
h_{22}(t) = <{\bf 2}|  \bigl(  {\cal  A}(t)  +  {\cal  B}(t)  \bigr) 
{\Theta}^{-1}|{\bf 2}>, \label{l17} \end{equation} Adding expression 
(\ref{l17})  to  its   complex   conjugate    one   gets   \cite{14} 
\begin{equation} {\rm Re} \; (h_{11}(t) - h_{22}(t)) = {\rm  Re}  \; 
<{\bf 2}| \bigl( {\cal A}(t) + {\cal B}(t) \bigr) {\Theta}^{-1}|{\bf 
2}>. \label{l18} \end{equation}

Note that  if  to  replace  the  the 
requirement (\ref{l7}) for  the  projector  $P$  (\ref{l3})  by  the 
following   one:   \begin{equation}   [P,H]   =    0,    \label{dd1} 
\end{equation} i.e., if to consider only stationary  states  instead 
of unstable states, then one immediately  obtains  from  (\ref{l11}) 
--- (\ref{l14}):  \begin{eqnarray}  {\cal  A}(t)  \;  &  =  &  \;  P  
[{\Theta},H] P,   \label{dd2}  \\  {\cal  B}(t)  \;  &  =  &  \;  0. 
\label{dd3} \end{eqnarray}

\section{The  case  of  conserved  CPT-- symmetry.}  

Let  us  assume  
that  condition  (\ref{l7})  holds  and 
\begin{equation} [\Theta,H] = 0. \label{l19} \end{equation} For  the 
stationary   states   (\ref{dd1}),   this   assumption,    relations 
(\ref{dd2}),  (\ref{dd3})  and  (\ref{l18})  yield  ${\rm   Re}   \; 
(h_{11}(t) - h_{22}(t)) = 0$.

Now  let  us  consider  the  case  of 
unstable states, i.e., states $|{\bf 1}>, |{\bf 2}>$, which lead  to 
such projection operator $P$ (\ref{l3})  that  condition  (\ref{l7}) 
holds. If in this case (\ref{l19})  also  holds  then  ${\cal  A}(t) 
\equiv 0$ and thus  $[ {\Theta}, H_{\parallel}(0) ]$ $ =  0$,  which 
is in agreement  with  an  earlier,  similar  result  \cite{11}.  In 
this case  we  have  ${\Theta}U(t)  =  U^{+}(t)\Theta$, which  gives 
${\Theta}U_{\parallel}(t)      =$      $U_{\parallel}^{+}(t)\Theta$,  
${\Theta}U_{\parallel}^{-1}(t)      =      (U_{\parallel}^{+}(t))^{- 
1}\Theta$, and \begin{equation} [\Theta, U(t)]  =  - 2i \bigl(  {\rm 
Im} \; U(t) \bigr) \Theta \label{l20} \end{equation}  This  relation 
leads to the following result in  the  case  of  the conserved CPT-- 
symmetry \begin{eqnarray} {\cal B}(t) & = & -2i P \Big{\{}  H  \:  - 
\: H_{\parallel}(t) \: P \Big{\}} \bigl( {\rm Im} \: U(t)  \bigr)  P 
\bigl(  U_{\parallel}^{+}(t)  {\bigr)}^{-1}\Theta  \label{l21}\\   & 
\equiv & -2i \Big{\{} PHQ \: - \:  V_{\parallel}(t)  \:  P  \Big{\}} 
\bigl(  {\rm  Im}  \:  U(t)  \bigr)  P  \bigl(  U_{\parallel}^{+}(t) 
{\bigr)}^{-1} \Theta  .  \label{l22}  \end{eqnarray}

The  simplest, 
nontrivial case is the case of $\dim {\cal H} = 3$: here $\Upsilon = 
{\{}1,2,3  {\}}$  and  (see  (\ref{l7c}))  \[  Q  =  |e_{3}><e_{3}|, 
\label{l23} \] and all operators acting in $\cal H$ can be  realized 
as $3 \times 3$ matrices. In this  representation:  \[  P  =  \left( 
\begin{array}{ccc} 1 & 0 & 0 \\ 0 & 1 & 0 \\ 0 & 0 &  0  \end{array} 
\right) , \; \; Q = \left( \begin{array}{ccc} 0 & 0 & 0 \\ 0 & 0 & 0 
\\ 0 & 0 & 1 \end{array} \right) , \label{l24}  \]  \[  H  =  \left( 
\begin{array}{ccc} H_{11} &  H_{12}  &  H_{13}  \\  H_{12}^{\ast}  & 
H_{22}  &  H_{23}  \\  H_{13}^{\ast}  &   H_{23}^{\ast}   &   H_{33} 
\end{array}   \right)   ,   \;    \;    V_{\parallel}    =    \left( 
\begin{array}{ccc} v_{11} & v_{12} & 0 \\ v_{21} & v_{22} & 0 \\ 0 & 
0  &  0  \end{array}  \right)  ,  \label{l25}  \]   (where   $v_{jk} 
\stackrel{\rm def}{=} <{\bf j}|V_{\parallel}(t) |{\bf k}>$; $j,k  \, 
= \, 1,2$) and therefore for   ${\{}   PHQ   -   V_{\parallel}(t)  P 
{\}}$    in   (\ref{l22})   we   obtain   \begin{equation}   PHQ   - 
V_{\parallel}(t)  P = \left( \begin{array}{ccc} - v_{11} & -  v_{12} 
& H_{13} \\ - v_{21} & - v_{22} & H_{23} \\ 0 & 0  &  0  \end{array} 
\right) . \label{l26} \end{equation}

Denoting $\bigl(  {\rm  Im}  \: 
U(t)   \bigr)   P    \bigl(    U_{\parallel}^{+}(t)    {\bigr)}^{-1} 
\stackrel{\rm def}{=} {\cal Z}(t)$ one finds  that  \begin{equation} 
{\cal Z}(t) = \left( \begin{array}{ccc}  z_{11}  &  z_{12}  &  0  \\ 
z_{21} & z_{22} & 0 \\ z_{31} & z_{32} &  0  \end{array}  \right)  , 
\label{l27}  \end{equation}  where  $z_{lm}  \stackrel{\rm   def}{=} 
<e_{l}|{\cal Z}|e_{m}>$, ($l,m \, = \, 1,2,3$).  It  is  clear  that 
${\cal Z}(0) = 0$ and ${\cal  Z}(t)  \neq  0$  for  $t  >  0$.  From 
(\ref{l26}) and ({\ref{l27}) it is seen that in the considered  case 
of $\dim {\cal H} = 3$, the operator ${\cal B}(t)$  (\ref{l22})  has 
the  form  \begin{equation}  {\cal  B}(t)   =   -2i   {\Big(}PHQ   - 
V_{\parallel}(t)P{\Big)}  {\cal  Z}(t)  \Theta  \equiv  -2i   \left( 
\begin{array}{ccc} b_{11} & b_{12} & 0 \\ b_{21} & b_{22} & 0 \\ 0 & 
0 & 0 \end{array} \right) \Theta ,  \label{l28}  \end{equation}  and 
that $b_{jk}(0) = 0$ and $b_{jk}(t) \neq 0$ for $t > 0$ in  case  of 
conserved CPT--symmetry (\ref{l19}). Thus, in our case, ${\cal B}(0) 
\equiv 0$ and ${\cal B}(t) \neq 0$ for $t > 0 $. Similar  conclusion 
holds in the case of $\dim {\cal H} > 3$.  Generally,  in  any  case 
${\cal B}(t > 0) \neq 0$.

Formulae (\ref{l21}), (\ref{l22}) and  the 
example considered above allow us to conclude that  $<{\bf  2}|{\cal 
B}(0){\Theta}^{-1} |{\bf 2}> = 0$ and ${\rm Re}<{\bf 2}|{\cal B}(t > 
0) {\Theta}^{-1}|{\bf 2}> \neq 0$, if condition  (\ref{l19})  holds. 
This means that in this case it  must  be  ${\rm  Re}(\,h_{11}(t)\,) 
\neq  {\rm  Re}(\,h_{22}(t)\,)$  for  $t  >  0$.  So,  there  is  no 
possibility for ${\rm Re}(h_{11})$ to equal ${\rm  Re}(h_{22})$  for 
$t> 0$  in  the  considered case  of  $P$  fulfiling  the  condition 
(\ref{l7}) (i.e.,  for  unstable  states)   when  CPT--symmetry   is 
conserved: It must be ${\rm Re}(h_{11}) \neq {\rm  Re}(h_{22})$  and 
thus  $h_{11}  \neq  h_{22}$  in   such  a  case.  \\    \hfill   \\

\section{Discussion.} 

Assuming that  condition  (\ref{l7})  for  $P$ 
holds one finds that the only possibility for Re$(h_{11} -  h_{22})$ 
to be equal zero appears if the nonzero contribution  of  \linebreak 
${\cal B}(t > 0){\Theta}^{-1}$ into Re$(h_{11}(t) - h_{22}(t))$   is  
compensated by a nonzero contribution of ${\cal  A}(t){\Theta}^{-1}$ 
--- see  (\ref{l18}).  It   can   be    observed    that  \linebreak 
Re$<{\bf   2}|{\cal   B}(t   >0){\Theta}^{-1}|{\bf   2}>   \neq   0$ 
irrespectively  of whether  $\Theta$ commutes with $H$ or  not,  but 
${\cal A}(t)  \neq 0$ and  $<{\bf  2}|{\cal  A}(t){\Theta}^{-1}|{\bf 
2}> \neq 0$ only appear if $[\Theta,H] \neq 0$. So,  all  the  above 
considerations lead to the following conclusions  for   the   matrix 
elements   $h_{jk}$   of    the  exact     effective     Hamiltonian 
$H_{\parallel}$ governing the time  evolution   in   neutral   kaons 
subspace: \\  \hfill  \\  {\noindent}{\it  Conclusions 
1}:  \\ 1.a) If ${\rm Re.}(\,h_{11}(t > 0)\,) = {\rm Re.}(\,h_{22}(t 
> 0)\,)$ then it  follows  that  $[\Theta,H]   \neq0$,  \\  1.b)  If 
$[\Theta,H] = 0$ then it follows  that  ${\rm  Re.}(\,h_{11}(t>0)\,) 
\neq    {\rm    Re.}(\,h_{22}(t>0)\,)$.    \\    1.c)    If    ${\rm 
Re.}(\,h_{11}(t>0)\,)  \neq  {\rm  Re.}(\,h_{22}(t>0)\,)$  then  the 
cases  $[\Theta,H]  \neq   0$  or  $[\Theta,H]   =  0$   are    both 
possible. \\ \hfill \\

One should remember  that  above  conclusions 
derived from relation (\ref{l18}) concern only  the  real  parts  of 
$h_{11}(t>0)$   and   $h_{22}(t>0)$.   Relations   (\ref{l16})   --- 
(\ref{l18})  give us  no information about the  imaginary  parts  of  
$h_{11}$  and  $h_{22}$. One cannot infer from   ({\ref{l18})   that  
$[\Theta   ,H]   =   0$  follows   ${\rm   Im}(h_{11})   \neq   {\rm 
Im}(h_{22})$.  The  case  when  $[  \Theta  ,H]  =0$  follows  ${\rm 
Re}(\,h_{11}(t>0)\,) \neq  {\rm  Re}  (\,h_{22}(t>0)\,)$  and  ${\rm 
Im}(h_{11})  =  {\rm  Im}(h_{22})$,   is   not   in  conflict   with 
relations (\ref{l16}) --- (\ref{l18}).  The equality of Im$(h_{11})$ 
and Im$(h_{22})$ need not imply the  equality  of  Re$(h_{11})$  and  
Re$(h_{22})$  and  vice  versa. This  means that   Bell--Steinberger  
relations  \cite{16} do not  contradict  relations  (\ref{l16})  --- 
(\ref{l18})  and Conclusions 1.a) --- 1.c) following from them: Bell  
and  Steinberger formulae lead to the equality  of Im$(h_{11})$  and  
Im$(h_{22})$  in  the  case  of  conserved  CPT--symmetry and do not 
concern  the   real   parts   of   diagonal   matrix   elements   of 
$H_{\parallel}$ and do not give relations between them.

Real  parts 
of diagonal matrix elements  of  the  mass  matrix  $H_{\parallel}$, 
$h_{11}$ and $h_{22}$, are considered in the literature as masses of 
unstable particles $|{\bf 1}>, |{\bf 2}>$ (eg., mesons ${\rm K}_{0}$ 
and 2${\overline{\rm K}}_{0}$). Conlusion 1.b) means that masses  of 
a decaying particle "1" and its antiparticle "2" should be different 
if CPT--symmetry is conserved in the system containig these unstable 
particles. In other words,  in  the  exact  theory  unstable  states 
$|{\bf 1}>, |{\bf 2}>$appear to be nondegenerate in  mass  if  CPT-- 
symmetry holds in the total system considered.  At  the  same  time, 
relations(\ref{dd1}) ---  (\ref{l19})  suggest  that  in  the  CPT-- 
invariant system masses of a given particle and its aniparticle  are 
equal (i.e., apear to be degenerate) only in the case of  stationary 
(stable) states $|{\bf 1}>,  |{\bf  2}>$.  The  case,  when  vectors 
$|{\bf  1}>,  |{\bf   2}>$   describe   pairs   of   particles   $p, 
\overline{p}$, or $e^{-}, e^{+}$, can be considered as an example of 
such states. All these conclusions contradict the standard result of 
the LOY and related approaches.

Results of  Sections  4  and  5  and 
Conclusions 1.a) --- 1.c) are not in conflict with such implications 
of the CPT--invariance as the equality of particle and  antiparticle 
decay rates. It can be easilly verified  that  assuming  (\ref{l19}) 
one gets (see (\ref{l4}) and (\ref{l9})  )  \begin{equation}  |<{\bf 
2}|U_{\parallel}(t)|{\bf 2}>|^{2} = |<{\bf  1}|U_{\parallel}(t)|{\bf 
1}>|^{2}. \label{l30} \end{equation} This last relation  means  that 
decay  laws,  and  thus  decay  rates,  of  particle  "1"  and   its 
antiparticle  "2"  are  equal.

The  consequences   (\ref{b8})   and 
(\ref{b9}) of the LOY theory are in conflict  with  the  results  of 
Sec. 4 and 5,  and   Conclusions  1.a)   --  1.c)  obtained  without 
approximations. From Conclusions 1.a) --- 1.c)  we  infer  that  for  
experimentally  measured  parameter  $\delta$  (\ref{r9})  following 
Conclusions  should  be   valid   \cite{14}:    \\   \hfill \\
{\noindent}{\it 
Conclusions 2}:\\ 2.a) If $\delta  =  0$  (or  ${\varepsilon}_{l}  = 
{\varepsilon}_{s}$) then it follows that $[\Theta,H]   \neq  0$.  \\ 
2.b) If $[\Theta,H] = 0$ then it follows that  $\delta  \neq 0$  (or 
${\varepsilon}_{l} \neq {\varepsilon}_{s}$). \\ 2.c) If $\delta \neq 
0$ ( or ${\varepsilon}_{l} \neq {\varepsilon}_{s}$) then  the  cases 
$[\Theta,H] \neq  0$  or $[\Theta,H]  =  0$  are   both  possible.\\  
\hfill \\

Properties of the real systems described in Conclusions  1  
and  2 are unobservable for  the  LOY  approximation.  In  order  to 
obtain  at least  an  estimation  for  effects  described  in  these 
Conclusions, matrix elements of $H_{\parallel}$ should be calculated 
much more exactly than it is  possible  within  the  LOY  theory.  A 
proposal  of more exact approximation is given in  \cite{9,10}  (see 
Appendix B). All  CP --  and CPT -- transformation properties of the 
effective  Hamiltonian  $H_{\parallel}$  calculated   within    this  
approximation are  consistent  with  similar   properties   of   the  
exact  effective Hamiltonian.  For  instance,  using  the  formalism 
mentioned  (and,  for  readers  convenience,  briefly  described  in 
Appendix B),  one  can   find$(h_{11}  -  h_{22})$  for  generalized 
Fridrichs--Lee model \cite{chiu}.  Assuming  CPT--invariance  (i.e., 
(\ref{l19}), it  has been  found  in  \cite{9}  (see formulae (152), 
(144) and (145) in \cite{9}) that \begin{eqnarray} 2h_{z} & \equiv & 
h_{11}  -  h_{22}  =  \lim_{t \rightarrow \infty} \Big( h_{11}(t)  - 
h_{22}(t)  \Big)   \label{FL1}   \\   &   \equiv   &    \frac{m_{21} 
{\Gamma}_{12} - m_{12}{\Gamma}_{21}}{|4m_{12}|} \Big( F_{0}(m_{0}  -  
\mu - |m_{12}|) - ( F_{0}(m_{0} - \mu + |m_{12}|)  \Big),  \nonumber 
\end{eqnarray}  where,   ${\Gamma}_{12},   {\Gamma}_{21}$   can   be 
identified with those apperaing  in   the  LOY  theory  (\ref{b3a}), 
(\ref{b5a}), $m_{0} \equiv H_{11} = H_{22}$ can be   considered   as  
kaon mass \cite{chiu}, $m_{jk} \equiv H_{jk} \, (j,k  =1,2)$,  $\mu$ 
can be treated as the mass of the decay  products  of   the  neutral  
kaon \cite{chiu}, and \begin{eqnarray*} F_{0}(m) & =  &  i  m^{-1/2} 
a_{1}(0),  \\   a_{1}(0)   &   \equiv   &   (m_{0}   -   \mu)^{1/2}, 
\end{eqnarray*} which, in  the  case  of  conserved   CPT--symmetry, 
lead  to   the following estimation for $|m_{12}|  \ll  (m_{0}-  \mu 
)$:  \begin{eqnarray}  h_{11}   -   h_{22}   &  \simeq  &  i  \frac{ 
m_{21}{\Gamma}_{12}  -  m_{12}{\Gamma}_{21}  }{4(m_{0}  -   \mu   )} 
\nonumber  \\  &  \equiv  &  \frac{  {\rm   Im}(m_{12}{\Gamma}_{21}) 
}{2(m_{0}- \mu )} \stackrel{\rm df}{=} 2h_{z}^{FL} \label{FL2}  \\ & 
\equiv &  {\rm  Re}(  2h_{z}^{FL}  ).  \nonumber  \end{eqnarray}  An 
equivalent  form  of  this  estimation   is   the   following   one: 
\begin{equation} {\rm Re} (2h_{z}^{FL})  = \frac{- {\rm Re}( m_{12}) 
\;    {\rm    Im}({\Gamma}_{12})    +     {\rm     Im}(m_{12})\;{\rm 
Re}({\Gamma}_{12}) }{2(m_{0} - \mu  )}.  \label{FL3}  \end{equation} 
Real properties of neutral K--complex enable  us  to  replace  $\tan 
{\phi}_{SW}$  (\ref{new30})  in  (\ref{new29})  by  "1"  \cite{data, 
dafne}:  $\tan  {\phi}_{SW}  \approx  1$,  and  to  use   assumption 
(\ref{new21})  in  formula  (\ref{new29})  for  Im$({\Gamma}_{12})$. 
Thefore keeping in mind relations (\ref{new27}),  (\ref{new29})  one 
can  conclude  that  the  contribution  Im$({\Gamma}_{12})$  in  the 
numerator of (\ref{FL3})  is neglegibly small in comparison with the 
contribution of Re$({\Gamma}_{12})$ in the case of neutral K--mesons 
considered. Finally, using (\ref{new27}), the estimation (\ref{FL3}) 
takes the following form: \begin{eqnarray} {\rm Re} (2h_{z}^{FL})  & 
= & {\rm Im}(m_{12})\; \frac{ {\gamma}_{s} - {\gamma}_{l} } {4(m_{0} 
- \mu  )}  \nonumber  \\  &  \approx  &  {\rm  Im}(m_{12})  \;\frac{ 
{\gamma}_{s} }{4(m_{0} -  \mu  )}.  \label{FL4}  \end{eqnarray}  For 
neutral  K--system,  to  evaluate   $2h_{z}^{FL}$   one   can   take 
${\tau}_{s} \simeq 0,89  \times  10^{-10}  {\rm  sec}$  \cite{data}. 
Hence ${\gamma}_{s} = \frac{\hbar}{{\tau}_{s}} \sim 7,4 \times 10^{- 
12} {\rm MeV}$ and (following \cite{chiu} ) $(m_{0} - \mu ) =  m_{K} 
- 2m_{\pi} \sim 200$ MeV \cite{chiu}.  Thus  $2h_{z}^{FL}\sim   0,93 
\times 10^{-14} {\rm Im.}(m_{12})\equiv 0,93  \times  10^{-14}  {\rm 
Im.}(H_{12})$.

Searching for another properties of  the Friedrichs-- 
Lee model it has been also found in \cite{9} that $h_{jk}(t)  \simeq 
h_{jk}$ practically for  $t  \geq  T_{as}  \simeq  \frac{10^{2}}{\pi 
(m_{0} - \mu - |m_{12}|)}$ (see \cite{9}, formula (153)). Therefore, 
within assumptions used above we obtain $T_{as} \sim 10^{-22}$  sec.

On the other hand, relations (\ref{FL4}) can be rewritten to  obtain 
an estimation for the matrix element $H_{12}$ of the  CPT--invariant 
Hamiltonian $H$: \begin{equation}  {\rm  Im}  (m_{12})  \equiv  {\rm 
Im}(H_{12}) \simeq 4 \frac{{\rm Re}(h_{11} - h_{22})}{{\gamma}_{s} - 
{\gamma}_{l}} (m_{0}  -  \mu  ).  \label{FL5}  \end{equation}  Using 
relation (\ref{new34}) one can express  Re.$(h_{11}  -  h_{22})$  in 
terms of physical parameters and then  inserting  into  the  formula 
(\ref{new34})  experimentally  obtained  present  data   for   these 
parameters one can obtain a numerical value of Re$(h_{11} - h_{22})$ 
\cite{data,  dafne}:  \begin{equation}  \frac{|{\rm   Re}(h_{11}   - 
h_{22})|    }{m_{K_{0}\,average}}     \equiv     \frac{|M_{11}     - 
M_{22}|}{m_{K_{0}\,average}} \sim 9 \times  10^{-19}  .  \label{FL6} 
\end{equation} This, together with the estimation $(m_{0}  -  \mu  ) 
\sim   200$   MeV   used   above,    leads    to    the    following 
relation\begin{equation} |{\rm Im}(H_{12})| \sim  72 \times 10^{-17} 
\frac{m_{K_{0}\,average}}{{\gamma}_{s}}  [{\rm  MeV}]  \simeq   9,73 
\times 10^{-5}m_{K_{0} \, average}. \label{FL7} \end{equation}  Thus 
the estimation for imaginary part of matrix element ${\rm  Im}H_{12} 
\equiv$ \linebreak ${\rm Im.}< K_{0}|H|{\overline K}_{0} >$  of  the 
CPT--invariant  Hamiltonian  $H$  has  been   found.   This   simple 
estimation should be fulfiled by every  CPT--invariant model of weak 
interactions, which is expected to give  a  correct  explanation  of 
properties of neutral K--mesons. Therefore relation (\ref{FL7})  can 
be  also  considered   a  criterion  for  selfconsistency  of  model 
Hamiltonians $H$ of interactions  leading  to  a  decay  process  of 
neutral kaons.

Note  that  considering  in  detail  the  generalized 
Fridrichs--Lee model one finds that  ${\Gamma}_{jk}  =  0,  \;  (j,k 
=1,2)$ for $m_{0} < \mu$, i.e., for  bound  states  \cite{chiu,  9}. 
This observation and relations (\ref{FL2}), (\ref{FL3})  imply  that 
for bound (stable) states Re$h_{z}^{FL} = 0$. So, if CPT--symetry is 
conserved in this model, then particle and antiparticle bound states 
remain to be also degenerate in mass beyond the  LOY  approximation, 
whereas unstable states (i.e.,  states  for  which  $m_{0}  >  \mu$) 
appear to be nondegenerate in mass in  this  model  if  CPT--symetry 
holds. These observations confirm our earlier conlusions implied  by 
properties (\ref{dd1}) ---  (\ref{dd3}).

In  a  general  case,  in  
contradistinction  to  the property (\ref{b8})  obtained within  the 
LOY  theory,  one  finds   for   diagonal   matrix    elements    of 
$H_{\parallel}$  calculated  within   the   approximation    briefly 
sketched in Appendix B that  in  CPT--invariant system  (see  (B13), 
(B15)) \begin{equation}  h_{11}^{\mit   \Theta}   \neq  h_{22}^{\mit  
\Theta}, \label{b10}  \end{equation}  where  $h_{jk}^{\mit  \Theta}$  
denotes  matrix  elements  of  $H_{\parallel}^{\mit   \Theta}$   and  
$H_{\parallel}^{\mit \Theta}$ is the  operator $H_{\parallel}$  when 
the property (\ref{l19}) occurs. (The other approximation  improving 
WW formulae for $h_{jk}$ and used  in  \cite{3}  lead  to  the  same 
result). Similarly to the case  of  Fridrichs--Lee  model,  assuming 
that,  \begin{equation}  |H_{12}|  \ll  |H_{0}   |   ,   \label{b11} 
\end{equation} where \begin{equation} H_{0}  =  \frac{1}{2}  (H_{11} 
+ H_{22}), \label{b12} \end{equation} ($H_{0} = H_{11} = H_{22}$  if 
(\ref{l19}) holds) we find that (see (B18), (B19) ) \begin{equation} 
2 h_{z}^{\mit \Theta} \equiv  h_{11}^{\mit  \Theta}  -  h_{22}^{\mit 
\Theta} \simeq H_{12} \frac{ \partial {\Sigma}_{21}  (x)  }{\partial 
x} \begin{array}[t]{l} \vline \, \\  \vline  \,  {\scriptstyle  x  = 
H_{0} } \end{array}  -  H_{21}  \frac{  \partial  {\Sigma}_{12}  (x) 
}{\partial  x}  \begin{array}[t]{l}   \vline   \,   \\   \vline   \, 
{\scriptstyle  x  =   H_{0}   }\end{array}   \neq   0,   \label{b13} 
\end{equation} ($2h_{z}^{\mit \Theta} = 0$ only if  $[{\cal  C}{\cal 
P}, H] = 0$). This means that  if  CPT--symmetry  is  conserved   in  
the  system considered and (\ref{b11})  holds  then  the   parameter  
$\delta$   (\ref{r9})     takes     the     form    \begin{equation} 
{\delta}^{\mit \Theta} \equiv \frac{ 2h_{z}^{\mit \Theta} }{ D^{\mit 
\Theta} }, \neq  0,  \label{b14}  \end{equation}  where  \[  D^{\mit 
\Theta} \simeq D^{LOY} - ( H_{12} + H_{21} a^{-1} ) (1 +  a)  \frac{ 
\partial {\Sigma}_{0} (x) }{\partial x}  \begin{array}[t]{l}  \vline 
\, \\ \vline \, {\scriptstyle x =  H_{0}  }  \end{array}  ,  \]  and 
$D^{LOY}  =  h_{12}^{LOY}  +  h_{21}^{LOY}  +  \Delta  {\mu}^{LOY}$, 
$\Delta  {\mu}^{LOY}   \stackrel{\rm   def}{=}   {\mu}_{s}^{LOY}   - 
{\mu}_{l}^{LOY}$,  ${\mu}_{s}^{LOY}$  and    ${\mu}_{l}^{LOY}$   are 
eigenvalues of $H_{LOY}$ for eigenstates $|K_{s}>$   and   $|K_{l}>$ 
respectively, ${\Sigma}_{0}(H_{0})$ is  defined  by  formula  (B14), 
\begin{equation}      a      \stackrel{\rm       def}{=}       \Big( 
\frac{h_{21}^{LOY}}{h_{12}^{LOY}}   {\Big)}^{1/2}   .    \label{b15} 
\end{equation}



Confronting relations (\ref{b8}) with  (\ref{l18}),  or   (\ref{b9}) 
with Conclusions 2, one should remember that, in fact, $H_{LOY}$ can 
be considered as the lowest, nontrivial order approximation  in  the 
perturbation  $H_{W}$:  All  the  terms  to   higher   orders   than 
$(H_{W})^{2}$ are neglected in $H_{LOY}$ \cite{1} ---  \cite{6}.  It 
is obvious that CPT-- and other transformation properties of such an 
approximate effective Hamiltonian and of the exact one need  not  be 
the same. Taking into account all the above, it seems that  for  the 
proper understanding of CPT--invariance tests  and  CPT--invariance, 
or possible CPT--violation phenomena it  is  necessary  to  consider 
higher order contributions into $H_{\parallel}$ than those contained  
in  $H_{LOY}$.

Effects described above are, probably, beyond today's 
experiments accuracy, nevertheless nobody can exclude that  accuracy 
of future experiments will be much higher and the result  Re$(h_{11} 
- h_{22}) \neq 0$ will be obtained and  then  the  question  how  to 
interpret such a result could arise. The LOY  theory  is  unable  to 
give a correct interpretation of such  a  hipothetical  experimental 
result. For the correct interpretation  of  such  a  result,  matrix 
elements of $H_{\parallel}$ should be calculated much  more  exactly 
than it is possible within the LOY approach. The  result  (\ref{b8}) 
of the LOY approximation is model independent whereas, in  the  more 
exact theory, the magnitude of Re.$(h_{11} -  h_{22})$   depends  on 
the model of interactions considered. So a new  possibility  of  the 
verification of models of weak  interactions  arises  (see  formulae 
(\ref{FL5}) --- (\ref{FL7})).

It also seems, that above results have 
some meaning when attempts  to  describe  possible  deviations  from 
conventional  quantum  mechanics  are   made   and   when   possible 
experimental tests of  such  a  phenomenon  and  CPT--invariance  in 
neutral kaons system are considered \cite{12,13}. In such a  case  a 
very important role is played by nonzero contributions to $(h_{11} - 
h_{22})$ \cite{12,13}: The correct description of  these  deviations 
and experiments mentioned is impossible without taking into  account 
results of this and above Sections 2, 4 and 5.

In the light  of  the 
above discussion it is clear that it will  be  essential   for   the  
result  of  experimental   tests   of   the  CPT--invariance  to  be 
$|h_{11}   -   h_{22}|   \ll    |2h_{z}^{FL}|$    (\ref{FL2}).    In 
contradistinction  to  the  standard,  conventional   interpretation  
\cite{2} --- \cite{6},  such results  will  prove  that  $[\Theta,H] 
\neq 0$ in neutral  kaons,  or  other  similar,  systems.  The  same 
conclusion will follow  from  the  result  $|h_{11}  -  h_{22}|  \gg 
|2h_{z}^{FL}|$. There  is  a  chance  for  the  tested  system  that 
$[\Theta ,H] = 0$ only if  the  experiment  shows  that  $(h_{11}  - 
h_{22}) \sim 2h_{z}^{FL}$. Such an interpretation follows  from  the 
results of Sections 4 and 5, Conclusions 1, and from  properties  of 
generalized Fridrichs--Lee model \cite{chiu,9}. In the general case, 
all above conclusions are valid  if  one  replaces  $h_{z}^{FL}$  by 
$h_{z}^{\mit  \Theta}$  (\ref{b13}).

Analogous  consequences   will 
follow from the  following  results  of  experi\-ments:  Results  $| 
\delta  |  \ll  |{\delta}^{\mit  \Theta}|$   or   $|\delta   |   \gg 
|{\delta}^{\mit \Theta}|$ (\ref{b14}) will prove that  CPT--symmetry 
is violated by interactions  causing  decays  of  ${\rm  K_{0}},{\rm 
{\overline{K}}_{0}}$ mesons or  similar  systems.  Only  the  result 
$\delta  \sim {\delta}^{\mit \Theta} \neq 0$  can be  considered  as 
the confirmation  of  CPT--invariance  of  the  tested  system.

The 
problem is whether the experimenter will be able to   perform  their 
experiments with  the  accuracy  guaranteeing  the  proper answer to 
the question  of  whether  $|h_{11}  -  h_{22}|  \ll  |2h_{z}^{FL}|$ 
($|h_{11} - h_{22}| \ll |2h_{z}^{\mit \Theta}|$) and $|\delta |  \ll 
|{\delta}^{\mit \Theta}|$ or  $|h_{11} - h_{22}| \gg  |2h_{z}^{FL}|$ 
($|h_{11} - h_{22}| \gg |2h_{z}^{\mit \Theta}|$) and $|\delta |  \gg 
|{\delta}^{\mit \Theta}|$.

From Conclusions 1 and 2 it follows  that 
only the interpretation of results Re.$(h_{11} - h_{22}) =  0$   and 
$\delta = 0$ is uncontrovertible. Therefore only such results can be 
understood independently of the model.

The proper interpretation  of  
the  results Re$(h_{11} - h_{22})  \neq  0$  and  $\delta  \neq   0$  
depends  on  the  model  calculations of  the quantity $(h_{11}(t) - 
h_{22}(t))$ or, which is equivalent,  on the  calculated  values  of  
matrix elements  of  type  $<{\bf  2}|  {\cal  A}(t)|{\bf  1}>$  and 
$<{\bf  2}|{\cal B}(t)|{\bf  1}>$.  This can not be performed within 
the LOY approach  and requires  more exact approximations. It  seems 
that the approximation described in \cite{3}, the one  described  in 
Appendix B and  exploited in \cite{8} --- \cite{10} may  be  a  more 
effective tool for  this  purpose.

Above considerations suggest that 
tests consisting of a comparison of the  equality   of   the   decay 
laws of ${\rm K}_{0}$ and   ${\overline{\rm  K}}_{0}$  mesons,  i.e. 
verifying the relation (\ref{l30}), seem to be the only   completely  
model  independent  tests  for  verifying   the  CPT--invariance  in 
such and similar systems.

Taking into  account  all  the  above,  it 
seems that all theories describing the   time   evolution   of   the 
neutral kaons and similar systems  by   means   of   the   effective 
Hamiltonian $H_{\parallel}$ governing their   time   evolution,   in 
which the CPT--invariance of the total system  leads to the property 
(\ref{b8}) for this $H_{\parallel}$, (such as  LOY  theory  \cite{1} 
--- \cite{5} based on  the  WW  approximation),are  unable  to  give 
the exact  and  correct  description  of  allaspects of the  effects 
connected  with  the  violation  or nonviolation   of    the    CP--    
and    especially CPT--symmetries. (It occurs  probably  because  of 
the  fact  that  such  theories  cannot  exactly  satisfy  unitarity 
\cite{kabir} and lead to incosistencies of  CPT--symetry  properties 
of the $H_{\parallel}$ and the total  Hamiltonians  $H$  \cite{is}). 
Also, it seems that results of the experiments with  neutral  kaons, 
etc., designed  and  carried out  on  the  basis   of   expectations  
of  theories  within    WW  approximation,  such  as  tests  of  CPT 
invariance (at least  results  of  those  in  which  CPT--invariance 
or  CPT--noninvariance  of   $H_{\parallel}$   generated   by   such 
invariance  properties  of $H$ were essential),  should  be  revised 
using other methods  than the WW approach. \\  \hfill \\  \hfill  \\

{\Large\bf Appendix A.} \\ 
The aim of this Appendix is to  calculate 
the commutator $[ \Theta ,H_{\parallel}(t)]$ discussed  in  Sec.   2  
and to  study  some of its applications. In order to  calulate  this 
commutator it is convenient to express $H_{\parallel}(t)$  by  means 
of  the  formula  (\ref{l7a1}),  and  then   to    use    assumption 
(\ref{l9}), the definition of $[U_{\parallel}(t)]^{-1}$  (\ref{l6}), 
property   (\ref{l6a})   and   the   the   following   one   $$    P 
[U_{\parallel}(t)]^{-1}  =   [U_{\parallel}(t)]^{-1}  P   \equiv   P 
[U_{\parallel}(t)]^{-1}P, \eqno(A1) $$ which is the  consequence  of 
(\ref{l6}) and (\ref{l6a}).

Let us consider a commutator $[\Theta, P 
[U_{\parallel}(t)]^{-1} ]$.  It  is  only  non\-tri\-vial  relation, 
necessary for the  calculation  of  $[  \Theta  ,H_{\parallel}(t)]$. 
Using  the  property   $U_{\parallel}(t)   =   PU_{\parallel}(t)   = 
U_{\parallel}(t) P = P U_{\parallel}(t) P$ and relations (\ref{l6}), 
(A1)  we  find  (here  the   assumption   (\ref{l9})   is   crucial) 
\begin{flushright}        \begin{eqnarray*}        [\Theta,        P 
[U_{\parallel}(t)]^{-1} ] &=& \Theta P [U_{\parallel}(t)]^{-1}  -  P 
[U_{\parallel}(t)]^{-1}     \Theta      \\     &=&     \Theta      P 
[U_{\parallel}(t)]^{-1} - P [U_{\parallel}(t)]^{-1} P \Theta  \\ &=& 
P   U_{\parallel}^{-1}   \Big(   U_{\parallel}   \Theta   -   \Theta 
U_{\parallel} \Big) U_{\parallel}^{-1}  \\ &=& - P  U_{\parallel}^{- 
1} [ \Theta , U_{\parallel} ] U_{\parallel}^{-1} \\ & \equiv &  -  P 
U_{\parallel}^{-1} P [  \Theta  ,  U  ]  P  U_{\parallel}^{-1}.   \,  
\makebox[2in][r]{  ({\em  A}2)  }  \end{eqnarray*}  \end{flushright}

Relation   (\ref{l9b}),   properties     (A1)     and     expression 
(\ref{l7a1})  lead  to  the  following  formulae  \begin{flushright} 
\begin{eqnarray*} [ \Theta ,  H_{\parallel}(t)  ]  &=&  [  \Theta  , 
PHUPU_{\parallel}^{-1} P]  \\ &=& [ \Theta , PH ] UPU_{\parallel}^{- 
1} + PH [ \Theta, UP U_{\parallel}^{-1} ] \\ &=& P [ \Theta , H ] UP 
U_{\parallel}^{-1}  \,  \makebox[3in][r]{ ({\em A}3) }   \\  &+&  PH 
\Big{\{} [ \Theta , UP ]  U_{\parallel}^{-1}  +  UP  [  \Theta  ,  P 
U_{\parallel}^{-1} ] \Big{\}}. \end{eqnarray*} \end{flushright}  All 
steps  in  the  above  formulae and in formulae leading to (A2) have  
been  performed  without changing the order of  operators  appearing 
in products of type ${\Theta} H, {\Theta} U(t)$,  etc..  By   virtue  
of  the  assumption  (\ref{l9})   only   the   order   of  operators 
$\Theta$ and $P$ in products ${\Theta}P$, etc., can be changed  when 
it is necessary.  Now,  defining  $$  {\cal  A}  (t)   \stackrel{\rm 
def}{=}  P [ \Theta , H  ]  UP  U_{\parallel}^{-1}  ,  \eqno(A4)  $$ 
(which equals (\ref{l11}) ) and taking into account  (A2),  one  can 
obtain formula (\ref{l10}) from (A3) \begin{eqnarray*}  [  \Theta  , 
H_{\parallel}(t) ] & \equiv & {\cal A}(t)  +  PH  [  \Theta  ,  U  ] 
PU_{\parallel}^{-1}  \\ &+& PHUP \Big{\{} - U_{\parallel}^{-1}  P  [ 
\Theta, U ] P U_{\parallel}^{-1} \Big{\}}  \\ & = &  {\cal  A}(t)  + 
{\cal B}(t) , \end{eqnarray*} where (see (\ref{l12})) $$ {\cal B}(t) 
= \Big{\{} PH - PHUP U_{\parallel}^{-1} P \Big{\}}  [  \Theta  ,  U] 
PU_{\parallel}^{-1}, $$ or (by means of (\ref{l7a1})) \[ {\cal B}(t) 
\equiv  \Big{\{}  PH  -  H_{\parallel}  P\Big{\}}  [  \Theta  ,   U] 
PU_{\parallel}^{-1}, \] (i.e.,  simply  (\ref{l13})  ),   and    due   
to   the   properties (\ref{l7a}),  (\ref{l7b})  \[  {\cal  B}(t)  = 
\Big{\{}   PHQ   -   V_{\parallel}   P\Big{\}}   [   \Theta   ,   U] 
PU_{\parallel}^{-1}  ,  \]  that  is  formula  (\ref{l14}).

Let  us 
consider  now  some  details  of  the  derivation  of  the  relation 
(\ref{l17}).  Taking into account properties  of  the  anti--unitary 
operator $\Theta$ and CPT--trans\-for\-ma\-tion properties of states 
$| K_{0}>, |{\overline K}_{0}> $, etc., (see Sec.  2),  without  any 
assumptions for the commutator $[\Theta , H]$, one can transform the 
matrix element $<{\bf  2}|\Theta  H_{\parallel}(t)  \Theta^{-1}|{\bf 
2}>$ appearing in (\ref{l17})  as  follows  \begin{eqnarray*}  <{\bf 
2}|\Theta  H_{\parallel}(t)  \Theta^{-1}|{\bf   2}>   &   \equiv   & 
<{\overline  K}_{0}|\Theta  H_{\parallel}(t)  \Theta^{-1}|{\overline 
K}_{0}> \\ &  \equiv  &  <{\Theta}  K_{0},  \Theta  H_{\parallel}(t) 
\Theta^{-1}  {\Theta}  K_{0}>  \\  &  =  &   <{\Theta}^{-1}   \Theta 
H_{\parallel}(t) \Theta^{-1} K_{0}, {\Theta}^{-1} {\Theta} K_{0}> \\ 
&  =  &  <H_{\parallel}(t)  K_{0},  K_{0}>  \\   &   =   &   <K_{0}, 
H_{\parallel}(t)   K_{0}>^{\ast}   \\   &   \equiv   &   <{\bf   1}| 
H_{\parallel}(t)|{\bf   1}   >^{\ast}    \equiv    h_{11}(t)^{\ast}. 
\end{eqnarray*} This last relation and the following consequence  of 
({\ref{l16}) \[ <{\bf  2}|  \Theta  H_{\parallel}(t)  {\Theta}^{-1}| 
{\bf 2}> - <{\bf 2}|  H_{\parallel}(t)|{\bf  2}>  \equiv  <{\bf  2}| 
({\cal  A}(t)  +  {\cal  B}(t)){\Theta}^{-1}|  {\bf  2}>,  \]  yield 
\begin{equation} h_{11}(t)^{\ast} - h_{22}(t)  =  <{\bf  2}|  \bigl( 
{\cal  A}(t)  +   {\cal   B}(t)   \bigr)   {\Theta}^{-1}|{\bf   2}>, 
\end{equation} i.e., the formula (\ref{l17}). \\  \hfill  \\  
\hfill \\  

{\Large\bf  Appendix  B.}  \\  
The  approximate   formulae   for 
$H_{\parallel}(t)$ have been  derived  in  \cite{9,10}   using   the  
Krolikowski--Rzewuski equation   for   the  projection  of  a  state 
vector \cite{7}, which results from  the  Schr\"{o}dinger   equation 
(\ref{l2}) for  the  total system under consideration, and,  in  the  
case  of initial conditions  of  the  type  (\ref{l2a}),  takes  the 
following form $$  (  i  \frac{\partial}{  {\partial}  t}  -  PHP  ) 
U_{\parallel}(t)   =   -  i   \int_{0}^{\infty}   K(t   -   \tau   ) 
U_{\parallel} ( \tau ) d \tau,  \eqno (B1) $$ where $  U_{\parallel} 
(0)  =  P$, $$ K(t)  =  {\mit  \Theta}  (t)  PHQ  \exp  (-itQHQ)QHP, 
\eqno (B2) $$ and ${\mit \Theta} (t)  =  { \{ } 1 \;{\rm for}  \;  t 
\geq 0, \; \; 0 \; {\rm for} \; t < 0 { \} }$ .

Taking into  account 
(\ref{l7b}) one finds  from  (\ref{l1}),  (\ref{l1a})  and  (B1)  $$ 
V_{\parallel} (t) U_{\parallel} (t) = - i  \int_{0}^{\infty}  K(t  - 
\tau ) U_{\parallel} ( \tau ) d \tau \stackrel{\rm def}{=} - iK \ast 
U_{\parallel} (t) . \eqno (B3) $$ (Here the asterix  $\ast$  denotes 
the convolution: $f  \ast g(t) = \int_{0}^{\infty}\, f(t - \tau ) g( 
\tau  ) \, d \tau$ ).\linebreak Next,  using  this  relation  and  a 
retarded Green's  operator  $G(t)$ for the equation (B1) $$ G(t) = - 
i {\mit  \Theta}  (t)  \exp  (-itPHP)P,  \eqno(B4)  $$  one  obtains 
\cite{9,10} $$ V_{\parallel}(t) \; U_{\parallel}(t) =  -  i  K  \ast 
\Big[ {\it 1} + \sum_{n = 1}^{\infty} (-i)^{n}L \ast \ldots  \ast  L 
\Big] \ast U_{\parallel}^{(0)} (t) , \eqno  (B5)  $$  where  $L$  is 
convoluted $n$ times, ${\it 1} \equiv {\it 1}(t) \equiv \delta (t)$, 
$$ L(t) = G \ast K(t), \eqno (B6) $$ and  $$  U_{\parallel}^{(0)}  = 
\exp (-itPHP) \; P \eqno (B7) $$ is a "free" solution of  Eq.  (B1). 
Of course, the  series (B5) is convergent if  \linebreak  $\parallel 
L(t) \parallel < 1$. If for every  $t  \geq  0$  $$  \parallel  L(t) 
\parallel \ll 1, \eqno (B8) $$ then, to the lowest order of  $L(t)$,  
one   finds   from  (B5)  \cite{9,10}  $$   V_{\parallel}(t)   \cong 
V_{\parallel}^{(1)} (t) \stackrel{\rm def}{=}  -i  \int_{0}^{\infty} 
K(t - \tau ) \exp {[} i ( t - \tau ) PHP {]} d \tau . \eqno (B9)  $$ 
In the case of (\ref{l3}) of the projector  $P$,  this   approximate 
formula for $V_{\parallel}(t)$ leads to the  following   expressions 
for the matrix elements $v_{jk}(t \rightarrow \infty ) \stackrel{\rm  
def}{=}  v_{jk}$  of \linebreak $V_{\parallel}(t  \rightarrow \infty 
) \stackrel{\rm def}{=} V_{\parallel}  \cong  V_{\parallel}^{(1)}  ( 
\infty )$ \cite{9,10}, \begin{flushright} \begin{eqnarray*} v_{j1} = 
& - & \frac{1}{2} \Big( 1 + \frac{H_{z}}{\kappa} \Big) {\Sigma}_{j1} 
(H_{0} + \kappa ) - \frac{1}{2} \Big( 1 - \frac{H_{z}}{\kappa} \Big) 
{\Sigma}_{j1} (H_{0} - \kappa )   \\ & - &  \frac{H_{21}}{2  \kappa} 
{\Sigma}_{j2}  (H_{0}  +  \kappa   )   +   \frac{H_{21}}{2   \kappa} 
{\Sigma}_{j2} (H_{0} - \kappa ) ,  \\  &  \,  &  \makebox[4.5in][r]{ 
({\em  B}10)  }  \\  v_{j2}  =  &  -  &  \frac{1}{2}   \Big(   1   - 
\frac{H_{z}}{\kappa}  \Big)  {\Sigma}_{j2}  (H_{0}  +  \kappa  )   - 
\frac{1}{2}  \Big(  1  +  \frac{H_{z}}{\kappa}  \Big)  {\Sigma}_{j2} 
(H_{0} - \kappa )   \\ & - & \frac{H_{12}}{2  \kappa}  {\Sigma}_{j1} 
(H_{0} + \kappa ) + \frac{H_{12}}{2 \kappa} {\Sigma}_{j1}  (H_{0}  - 
\kappa ) , \end{eqnarray*}   \end{flushright} where $j,k = 1,2$,  $$ 
H_{z} = \frac{1}{2} ( H_{11} - H_{22} ) , \eqno (B11) $$ and $H_{0}$ 
is defined by (\ref{b12}), $$ \kappa = ( |H_{12} |^{2}  +  H_{z}^{2} 
)^{1/2} . \eqno (B12) $$ Hence, by (\ref{l7b}) $$ h_{jk} = H_{jk}  + 
v_{jk} . \eqno (B13) $$ These formulae for  $v_{jk}$  and  thus  for 
$h_{jk}$  have  been derived without assuming any symmetries of  the 
type  CP--,  T--,  or CPT--symmetry  for  the  total  Hamiltonian  H   
of   the   system considered. It should also be emphasized that  all  
components  of the expressions (B10)  have  the  same   order   with 
respect  to $\Sigma ( \varepsilon  )$.

In  the  case  of  preserved  
CPT--symmetry  (\ref{l19}),   one  finds  $H_{11}  =  H_{22}$  which 
implies that $\kappa \equiv |H_{12} |$, $H_{z} \equiv 0$ and  $H_{0} 
\equiv H_{11} \equiv H_{22}$, and  \cite{9,10}  $$  {\Sigma}_{11}  ( 
\varepsilon  =  {\varepsilon}^{\ast}  )   \equiv   {\Sigma}_{22}   ( 
\varepsilon   =   {\varepsilon}^{\ast}   )   \stackrel{\rm   def}{=} 
{\Sigma}_{0} ( \varepsilon = {\varepsilon}^{\ast} ) . \eqno (B14) $$ 
Therefore  matrix  elements   $v_{jk}^{\mit  \Theta}$   of  operator 
$V_{\parallel}^{\mit \Theta}$ ($V_{\parallel}^{\mit \Theta}$ denotes 
$V_{\parallel}$ when (\ref{l19}) occurs)  take  the  following  form 
\begin{eqnarray*} v_{j1}^{\mit \Theta} = & - & \frac{1}{2} {\Big\{ } 
{\Sigma}_{j1} (H_{0} + | H_{12} |) + {\Sigma}_{j1} (H_{0} - | H_{12} 
|)   \\ & +  &  \frac{H_{21}}{|H_{12}|}  {\Sigma}_{j2}  (H_{0}  +  | 
H_{12} |) - \frac{H_{21}}{|H_{12}|} {\Sigma}_{j2} (H_{0} - |  H_{12} 
|) {\Big\} } , \end{eqnarray*}  \hfill  (B15)  \\  \begin{eqnarray*} 
v_{j2}^{\mit \Theta} = & - &  \frac{1}{2}  {\Big\{  }  {\Sigma}_{j2} 
(H_{0} + | H_{12} |) + {\Sigma}_{j2} (H_{0} - | H_{12} |)   \\ & + & 
\frac{H_{12}}{|H_{12}|}  {\Sigma}_{j1}  (H_{0}  +  |  H_{12}  |)   - 
\frac{H_{12}}{|H_{12}|} {\Sigma}_{j1} (H_{0} - | H_{12}  |)  {\Big\} 
}, \end{eqnarray*}

Assuming (\ref{b11}),  we  find  $$  v_{j1}^{\mit  
\Theta} \simeq - {\Sigma}_{j1} (H_{0} )  -  H_{21}  \frac{  \partial 
{\Sigma}_{j2} (x) }{\partial x}  \begin{array}[t]{l}  \vline  \,  \\ 
\vline \, {\scriptstyle x = H_{0} } \end{array} , \eqno (B16) $$  $$ 
v_{j2}^{\mit  \Theta} \simeq  -  {\Sigma}_{j2}  (H_{0}  )  -  H_{12} 
\frac{ \partial {\Sigma}_{j1} (x) }{\partial x}  \begin{array}[t]{l} 
\vline \, \\ \vline \, {\scriptstyle x = H_{0} } \end{array} , \eqno 
(B17) $$ where $j = 1,2$.  One  should  stress   that   due   to   a 
presence of resonance terms,  derivatives  $\frac{\partial}{\partial 
x} {\Sigma}_{jk} (x)$ need not  be  small   and   neither  need  the  
products $H_{jk} \frac{\partial}{\partial x} {\Sigma}_{jk}  (x)$  in  
(B16).

Finally, assuming that (\ref{b11})) holds and using relations 
(B16),   (B13) and the expression (\ref{b5}),  we  obtain  for   the  
CPT--invariant system $$ h_{j1}^{\mit  \Theta}  \simeq   h_{j1}^{\rm 
LOY} -  H_{21}  \frac{  \partial  {\Sigma}_{j2}  (x)  }{\partial  x} 
\begin{array}[t]{l} \vline \, \\ \vline \, {\scriptstyle x = H_{0} } 
\end{array}  ,  \eqno  (B18)  $$  $$  h_{j2}^{\mit   \Theta}  \simeq  
h_{j2}^{\rm  LOY}  -  H_{12}  \frac{  \partial   {\Sigma}_{j1}   (x) 
}{\partial  x}  \begin{array}[t]{l}   \vline   \,   \\   \vline   \, 
{\scriptstyle x = H_{0} } \end{array} , \eqno (B19) $$  where  $j  = 
1,2$. From these formulae we  conclude  that, e.g.,  the  difference  
between   diagonal    matrix    elements    of  $H_{\parallel}^{\mit 
\Theta}$, which  plays   an   important   role  in  designing  CPT-- 
invariance  tests   for   the   neutral    kaons    system,   equals 
(\ref{b13}). Analogously, to the lowest order of  $|H_{12}  |$,  for 
eigenvalues ${\mu}_{l}, {\mu}_{s}$ (\ref{r5}) of $H_{\parallel}$, we 
obtain \cite{10} \begin{eqnarray*}  {\mu}_{s}^{\mit  \Theta}  \simeq 
{\mu}_{s}^{LOY} - \frac{1}{2} &  {\Big[  }  &  H_{12}  \Big(  \frac{ 
\partial {\Sigma}_{21} (x) }{\partial x} \begin{array}[t]{l}  \vline 
\, \\ \vline \, {\scriptstyle x = H_{0} }  \end{array}  +  a  \frac{ 
\partial {\Sigma}_{0} (x) }{\partial x}  \begin{array}[t]{l}  \vline 
\, \\ \vline \, {\scriptstyle x = H_{0} } \end{array} \Big)  \\ &  + 
& H_{21} \Big(  \frac{  \partial  {\Sigma}_{12}  (x)  }{\partial  x} 
\begin{array}[t]{l} \vline \, \\  \vline \, {\scriptstyle x =  H_{0} 
} \end{array} + a^{-1} \frac{ \partial {\Sigma}_{0}  (x)  }{\partial 
x} \begin{array}[t]{l} \vline \, \\  \vline  \,  {\scriptstyle  x  = 
H_{0} } \end{array} \Big)  {\Big] } , \end{eqnarray*}  \hfill  (B20) 
\\ \begin{eqnarray*} {\mu}_{l}^{\mit \Theta} \simeq  {\mu}_{l}^{LOY} 
-  \frac{1}{2}  &  {\Big[  }  &   H_{12}   \Big(   \frac{   \partial 
{\Sigma}_{21} (x) }{\partial x}  \begin{array}[t]{l}  \vline  \,  \\ 
\vline \, {\scriptstyle x = H_{0} } \end{array} - a \frac{  \partial 
{\Sigma}_{0} (x) }{\partial  x}  \begin{array}[t]{l}  \vline  \,  \\ 
\vline \, {\scriptstyle x = H_{0} } \end{array}  \Big)   \\  &  +  & 
H_{21}  \Big(  \frac{  \partial  {\Sigma}_{12}  (x)  }{\partial   x} 
\begin{array}[t]{l} \vline \, \\ \vline \, {\scriptstyle x = H_{0} } 
\end{array} - a^{-1} \frac{ \partial {\Sigma}_{0} (x) }{\partial  x} 
\begin{array}[t]{l} \vline \, \\ \vline \, {\scriptstyle x = H_{0} } 
\end{array} \Big)  {\Big] } , \end{eqnarray*} where $a$  is  defined 
by    (\ref{b15}).

\end{document}